\documentclass[twocolumn]{aastex63}

\newcommand{\xie}{\color{black}}
\newcommand{\jiwei}{\color{black}}
\newcommand{\chen}{\color{black}}
\newcommand{\respondtoWang}{\color{black}}
\newcommand{\respondtoxiang}{\color{black}}
\newcommand{\respondtohuang}{\color{black}}
\newcommand{\respondtoZheng}{\color{black}}

\usepackage{amsmath}
\usepackage{amssymb}
\usepackage{color}
\usepackage{url}
\usepackage{ulem}
\usepackage{multirow}
\usepackage{graphics}
\usepackage{longtable}
\usepackage{graphicx}
\usepackage{appendix}

\begin{document}

\title{Planets Across Space and Time (PAST). \uppercase\expandafter{\romannumeral1}. Characterizing the Memberships of Galactic Components and Stellar Ages: Revisiting the Kinematic Methods and Applying to Planet Host Stars}

\correspondingauthor{Ji-Wei Xie}
\email{jwxie@nju.edu.cn}


\author{Di-Chang Chen}
\affiliation{School of Astronomy and Space Science, Nanjing University, Nanjing 210023, China}
\affiliation{Key Laboratory of Modern Astronomy and Astrophysics, Ministry of Education, Nanjing 210023, China}

\author{Ji-Wei Xie}
\affiliation{School of Astronomy and Space Science, Nanjing University, Nanjing 210023, China}
\affiliation{Key Laboratory of Modern Astronomy and Astrophysics, Ministry of Education, Nanjing 210023, China}

\author{Ji-Lin Zhou}
\affiliation{School of Astronomy and Space Science, Nanjing University, Nanjing 210023, China}
\affiliation{Key Laboratory of Modern Astronomy and Astrophysics, Ministry of Education, Nanjing 210023, China}

\author{Su-Bo Dong}
\affiliation{Kavli Institute for Astronomy and Astrophysics, Peking University, Beijing 100871, China}

\author{Chao Liu}
\affiliation{Key Lab of Space Astronomy and Technology, National Astronomical Observatories, CAS, 100101, China}
\affiliation{University of Chinese Academy of Sciences, Beijing, 100049, China.}

\author{Hai-Feng Wang}
\affiliation{South-Western Institute for Astronomy Research, Yunnan University, Kunming, 650500, China; LAMOST Fellow}

\author{Mao-Sheng Xiang}
\affiliation{National Astronomical Observatories, Chinese Academy of Sciences, Beijing 100012, China}
\affiliation{Max-Planck Institute for Astronomy, Königstuhl 17, D-69117 Heidelberg, Germany}

\author{Yang Huang}
\affiliation{South-Western Institute for Astronomy Research, Yunnan University, Kunming, 650500, China}

\author{Ali Luo}
\affiliation{National Astronomical Observatories, Chinese Academy of Sciences, Beijing 100012, China}

\author{Zheng Zheng}
\affiliation{Department of Physics and Astronomy, University of Utah, Salt Lake City, UT 84112}

\begin{abstract}
Over 4,000 exoplanets have been identified and thousands of candidates are to be confirmed.
The relations between the characteristics of these planetary systems and the kinematics, {\respondtoxiang Galactic} components, and ages of their host stars have yet to be well explored. 
Aiming to addressing these questions, we conduct a research project, dubbed as PAST (Planets Across Space and Time).
To do this, one of the key steps is to accurately characterize the planet host stars.
In this paper, the Paper \uppercase\expandafter{\romannumeral1} of the PAST series, {\chen we revisit the  kinematic method for classification of Galactic components and extend the applicable range of velocity ellipsoid from $\sim 100$ pc to $\sim 1,500$ pc from the sun  in order to cover most known planet hosts.
Furthermore, we revisit the Age-Velocity dispersion Relation (AVR), which allows us to derive kinematic age with a typical uncertainty of 10-20\% for an ensemble of stars.
Applying the above revised  methods,} we present a catalog of kinematic properties (i.e. Galactic positions, velocities, the relative membership probabilities among the thin disk, thick disk, Hercules stream, and the halo) as well as other basic stellar parameters for {\chen 2,174} host stars of {\chen 2,872} planets by combining data from Gaia, LAMOST, APOGEE, RAVE, and the NASA exoplanet archive. 
{\chen The revised {\jiwei kinematic method and AVR} as well as the stellar catalog of kinematic properties and ages lay foundation for future studies on exoplanets from two dimensions of space and time in the Galactic context.}

\end{abstract}

 
 \section{introduction}
\begin{figure}[!ht]
\centering
\includegraphics[width=0.5\textwidth]{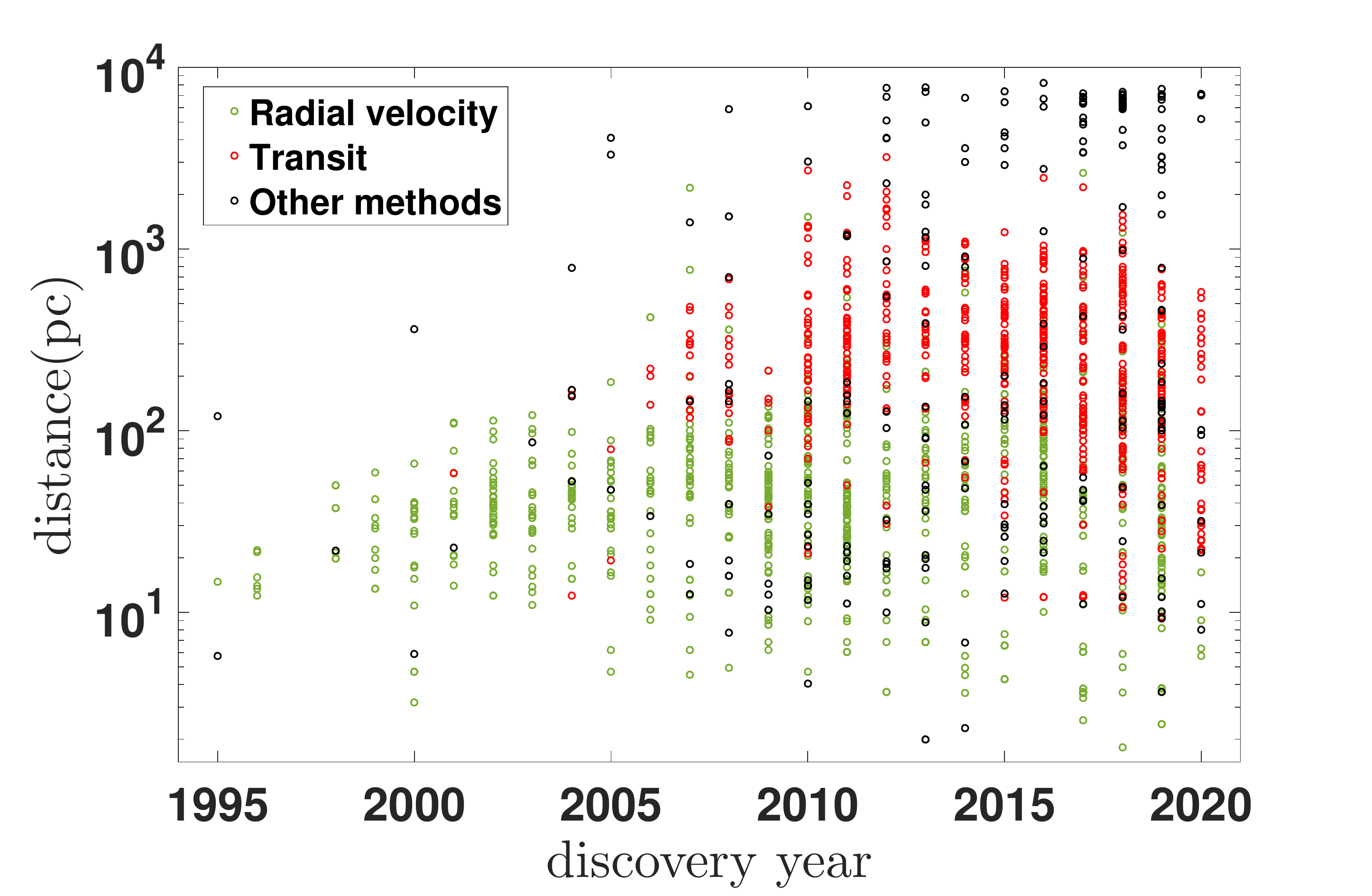}
\caption{The distance of planet host stars to the Sun vs. the year of discovery.
Planets discovered by different methods are plotted in different colours. 
\label{figeudistanceyear}}
\end{figure}

It has been a quarter century since the discovery of the first exoplanet. 
To date, over 4,000 exoplanets have been discovered and thousands of candidates are yet to be confirmed \citep[NASA Exoplanet Archive, EA hereafter;][] {2013PASP..125..989A}. 
There is a clear trend (shown in Figure \ref{figeudistanceyear}, data from http://exoplanet.eu) that our knowledge of exoplanets is expanding in the Galaxy.
Before 2005, most known exoplanets were confined in the solar neighborhood with distance less than $\sim$ 100-200 pc.
Now, the map of exoplanets is much wider with a large range of distance up to $\sim $ 10,000 pc.
Therefore, people began to study exoplanets in the context of the Galaxy.
For example, there were continuous discussions on how to define the {\respondtoxiang Galactic} habitable zone \citep{2001Icar..152..185G,2004Sci...303...59L,2006IJAsB...5..325S,2011AsBio..11..855G,2013AsBio..13..491J,2017NatSR...716626B,2019MNRAS.490..408S}, researches on the Galactic distribution of planets as a function of distance/population \citep[e.g.,][]{2017AJ....154..210Z}, and studies on whether planet occurrence rate depends on the {\respondtoxiang Galactic} velocity
\citep[e.g.,][]{2019MNRAS.489.2505M, 2019AJ....158...61B}.

One of fundamental questions in studying exoplanets in the {\respondtoxiang Galactic} context is: what are the differences in the properties of planetary systems at different positions in the Galaxy with different ages?
The answer of this question will provide insights on the formation and evolution of the ubiquitous and diverse exoplanets {\respondtoWang in different {\respondtoxiang Galactic} environments}.
Aiming to addressing the question, in a series of papers from here on, we conduct statistical studies of planets at different positions in the Galaxy with different ages, a project that we dub as PAST (Planets Across Space and Time).

To this end, the first step is to figure out where the exoplanets are in the Galaxy.
Specifically, for a given exoplanet host star, we would like to know which {\respondtoxiang Galactic} component (i.e., the thin disk, the thick disk or the halo) it belongs to.
One of well-established methods to distinguish these {\respondtoxiang Galactic} components is the kinematic approach as different components generally have different kinematic characteristics.
For example, the thin disk has a smaller vertical scale-height \citep{2012ApJ...753..148B,2018MNRAS.475.3633W},
but the thick disk is generally kinematically hotter with larger velocity dispersions \citep{1989ARA&A..27..555G,2003yCat..73400304R,2003A&A...410..527B,2014A&A...562A..71B,2018MNRAS.478.4513B}. 
By comparing the kinematic properties of a given star to the typical kinematic characteristics of a Galacic component, one may calculate the likelihood that the star belongs to this component \citep[e.g.][]{2003A&A...410..527B}. 
{\respondtoWang However, the kinematic characteristics of this method were obtained with data in the Solar neighborhood within $\sim$ 100 pc \citep{2003A&A...410..527B,2014A&A...562A..71B}, and thus limiting this kinematic method to a relatively small range of area.}
Thanks to the recent large scale star surveys  both from space \citep[e.g., Gaia, ][]{2016A&A...595A...1G,2018A&A...616A...1G,2018A&A...616A..11G} and ground \citep[e.g., LAMOST, ][]{1996ApOpt..35.5155W,2004ChJAA...4....1S,2012RAA....12.1197C,2012RAA....12..723Z,2012RAA....12.1243L}, we are now allowed to extend the kinematic method to beyond 1,000 pc in order to characterize the majority of exoplanet host stars. 

The second step is to obtain the ages of exoplanet host stars since most of exoplanet host stars have no (accurate) age estimates. 
Stellar ages can hardly be measured but only be inferred or estimated indirectly through a number of techniques, which have their own strength and weakness  \citep{2010ARA&A..48..581S}.  
For example, the widely-used isochrone placement method is applicable for estimating ages of a large range of stars, but it usually suffers from relatively large uncertainty ($\sim50\%$ typically) for main sequence stars, which are the bulk of exoplanet hosts \citep[e.g.][]{2020AJ....159..280B}. 
Asteroseismology is significantly better than any other age-dating method, which can deliver age estimates for individual stars with uncertainties of $\sim 10\%-20\%$ \citep[e.g.][]{2011ApJ...730...63G,2014ApJS..210....1C}.
However, this method requires observation with sufficiently accurate, high-cadence photometric measurements and it can only be applicable for stars with a limited range of spectral types that exhibit prominent oscillations.
In addition, the carbon and nitrogen abundances have been suggested to be age indicators, but it is usually applicable for giant stars, and  {\respondtoxiang the reported age has achieved a precision of $\sim$ 20\%-30\%
\citep{2016MNRAS.456.3655M,2016ApJ...823..114N,2017ApJ...841...40H,2018MNRAS.475.3633W}.}

Stellar ages can also be estimated statistically from some empirical relationships.
It has been known for decades that older stars have larger velocity dispersion, the so called Age-Velocity dispersion Relation (AVR) \citep{1946ApJ...104...12S,1950AZh....27...41P,1977A&A....60..263W,2009A&A...501..941H}. 
To derive age from AVR, one generally just needs the stellar kinematics, and thus the age is also called kinematic age.
Unlike the methods mentioned above, the kinematic method is only applicable to ensembles (not individual) of stars.
Nevertheless, kinematic age is still meaningful from a statistical view given the fast rise of exoplanet population.
Furthermore, the strength of kinematic age is that it uses only the 3D space motions (i.e., astrometry and radial velocities) without involving stellar evolutionary model, and thus it can apply to  stars in a large range of various parameters \citep{2010ARA&A..48..581S}.
In recent years, the kinematic method has ushered in a major opportunity, thanks to the high-quality astrometry and radial velocity observations for millions of stars (including thousands of exoplanet hosts) provided mainly by the Gaia and LAMOST.  

{In this paper, we revisit the methods to characterize stellar kinematic properties and apply them to over 2,000 exoplanet host stars based on their astrometry and radial velocities provided mainly by Gaia and LAMOST. 
Specifically, {\jiwei in section \ref{sec.meth.classify},  we revisit the kinematic method to identify Galactic components (e.g., thin/thick disk).
In section \ref{sec.meth.avr}, we revise the AVR to derive kinematic ages.
Applying the revised kinematic method and AVR, in section \ref{sec.res}, we present a catalog of kinematic properties for 2,174 planet host stars and conduct some analyses.}
In section \ref{sec.dis}, we discuss our results and some future prospects.
In section \ref{sec.guide}, we provide some importing guidelines, cautions, and limitations to utilize the kinematic methods and planet host catalog.
Finally, we summarize in section \ref{sec.sum}. }

\section{\chen Revisiting the Kinematic  Method to Classify the Galactic Components}
\label{sec.meth.classify}
{\jiwei In this section, we revisit the kinematic method to classify stars into different Galactic components (e.g., thin/thick disks). 
The key is revising the characteristic kinematic parameters (section \ref{sec.meth.rev}), an extension from the solar neighborhood ($\sim 100$ pc) to $\sim 1,500$ pc in order to cover most planet hosts as shown in Figure \ref{figeudistanceyear}.}


\subsection{Space Velocities and {\respondtoxiang Galactic} Orbits}
\label{sec.meth.space}
We calculated the 3D Galactocentric cylindrical coordinates $(R, \theta, Z)$ by adopting a location of the Sun of $R_\odot$= 8.34 kpc \citep{2014ApJ...783..130R} and $Z_\odot$ = 27 pc \citep{2001ApJ...553..184C}. 
The {\respondtoxiang Galactic} rectangular velocities relative to the Sun $(U, V, W)$ and their errors were calculated by the right-handed coordinate system based on the formulae and matrix equations presented in  \cite{1987AJ.....93..864J}.  
{\xie Here,} $U$ is positive when pointing {\xie to} the direction of the {\respondtoxiang Galactic} center, $V$ is positive along the direction of the Sun {\xie orbiting around the {\respondtoxiang Galactic} center}, and $W$ is positive when pointing towards the North {\respondtoxiang Galactic} Pole. 
Cylindrical velocities $V_R$, $V_\theta$, and $V_Z$ are defined as positive with increasing $R$, $\theta$, and $Z$, with the latter towards the North {\respondtoxiang Galactic} Pole. 
To obtain the {\respondtoxiang Galactic} rectangular velocities relative to  the local standard of rest (LSR) $(U_{\rm LSR}, V_{\rm LSR}, W_{\rm LSR})$, we adopted the solar peculiar motion [$U_\odot$, $V_\odot$, $W_\odot$] = [9.58, 10.52, 7.01] $\rm km \ s^{-1}$  \citep{2015ApJ...809..145T}. 



\subsection{Classification of {\respondtoxiang Galactic} Components}
\label{sec.meth.class}

We adopted the widely-used kinematic approach {\xie as in \cite{2003A&A...410..527B,2014A&A...562A..71B} to classify the stars in our sample into different {\respondtoxiang Galactic} components, e.g.,  thin and thick disk stars.}
This method assumes that the {\respondtoxiang Galactic} velocities ($U_{\rm LSR}, V_{\rm LSR}, W_{\rm LSR}$) in different components (the thin disk, the thick disk, the halo, and the Hercules stream) follow {\xie a multi-dimensional} Gaussian distribution as  
\begin{equation}
\begin{aligned}
&f (U,V,W) = k \times {\rm exp} \\
&\left(-\frac{(U_{\rm LSR}-U_{\rm asym})^2}{2{\sigma_U}^2}
-\frac{(V_{\rm LSR}-V_{\rm asym})^2}{2{\sigma_V}^2}
-\frac{W_{\rm LSR}^2}{2{\sigma_W}^2}\right),
\label{fUVW}
\end{aligned}
\end{equation}
where the normalization coefficient 
\begin{equation}
k = \frac{1}{(2\pi)^{3/2}\sigma_U\sigma_V\sigma_W}.
\label{kUVW}
\end{equation}
Here, $\sigma_U$, $\sigma_V$, and $\sigma_W$ are the characteristic velocity dispersions, and $V_{\rm asym}$ and $U_{\rm asym}$ are the asymmetric drifts. 

{\xie For $V_{\rm asym}$,  following \cite{2008gady.book.....B} we adopted 
\begin{equation}
V_{\rm asym} = \bar{V_{\theta}}-V_c,
\label{kUVW}
\end{equation}
where $V_c$ is the circular speed of LSR as 238 $\rm km \ s^{-1}$  \citep{2012MNRAS.427..274S} and $\bar{V_{\theta}}$ is the mean value of azimuthal velocities for a given component.}
{\xie For $U_{\rm asym}$, we adopted $U_{\rm asym}=0$ for the disk and halo components and $U_{\rm asym}=-40 \rm km \ s^{-1}$ for Hercules stream component \citep{2008gady.book.....B,2014A&A...562A..71B}.}

The relative probabilities {\xie between two different components, i.e.,  the thick-disk-to-thin-disk $(TD/D)$, thick-disk to halo $(TD/H)$, the Hercules-to-thin-disk $(Herc/D)$, and {\respondtoxiang the Hercules-to-thick-disk $(Herc/TD)$} can be calculated as }
\begin{alignat}{2}
 \frac{TD}{D} &=\frac{X_{\rm TD}}{X_{\rm D}} \cdot \frac{f_{\rm TD}}{f_{\rm D}},  &\quad \frac{TD}{H} &=\frac{X_{\rm TD}}{X_{\rm H}} \cdot \frac{f_{\rm TD}}{f_{\rm H}}, \\ 
 \frac{Herc}{D} &=\frac{X_{\rm Herc}}{X_{\rm D}} \cdot \frac{f_{\rm Herc}}{f_{\rm D}},  &\quad \frac{Herc}{TD} &=\frac{X_{\rm Herc}}{X_{\rm TD}} \cdot \frac{f_{\rm Herc}}{f_{\rm TD}},
\end{alignat}
\label{eqTDD}
where X is the fraction of stars for {\xie a given component}. 

{\xie Then for stars in our planet host sample, we calculated their above probabilities, and classified them into different {\respondtoxiang Galactic} components by adopting the same criteria as in \cite{2014A&A...562A..71B}, which are (1) thin disk: $TD/D<0.5 \,\,\&\,\, Herc/D<0.5$, (2) thick disk: $TD/D>2 \,\,\&\,\, TD/H>1 \,\,\&\,\, Herc/TD<0.5$, (3) halo: 
$TD/D>2 \,\,\&\,\, TD/H<1 \,\,\&\,\, Herc/TD<0.5$, and (4) Hercules:
$Herc/D>1 \,\,\&\,\, Herc/TD>1$.
}

\subsection{Revision of Characteristic Kinematic Parameters}
\label{sec.meth.rev}
{\xie One of the important steps during the above classification procedure is to obtain the characteristic kinematic parameters for each {\respondtoxiang Galactic} component, i.e., $\sigma_U$, $\sigma_V$, $\sigma_W$, $U_{\rm asym}$, $V_{\rm asym}$, and $X$.
In the solar neighbourhood within $\sim 100$ pc from the Sun, these parameters are available from \cite{2014A&A...562A..71B}, which are list in Table \ref{tab:dispersionsSN} here.}
However, {\xie as shown in Figure 1, the stars in our sample are} located in a much wider zone (up to several kpc from the Sun). 
It has been found that the velocity ellipsoids change with the {\respondtoxiang Galactic} position \citep{2013MNRAS.436..101W}.
Therefore, the values of these characteristic kinematic parameters for each component should be revised {\xie and extended (section \ref{sec.meth.rev}) so that they are  applicable for stars in a larger range of $Z$ (e.g., $|Z| < 1.5$ kpc)  and R ($7.5 < R < 10.0$ kpc).}

\begin{table}[!t]
\centering
\caption{
\label{tab:dispersionsSN}
        Characteristics for stellar components 
        in the Solar neighbourhood from \cite{2014A&A...562A..71B}.$^{\dagger}$
        }
\begin{tabular}{lcccccc}
\hline \hline\noalign{\smallskip}
        & $\sigma_{\rm U}$
        & $\sigma_{\rm V}$
        & $\sigma_{\rm W}$
        & $U_{\rm asym}$
        & $V_{\rm asym}$ 
        & $X$  \\
\noalign{\smallskip}
        & \multicolumn{5}{c}{-----------~~[km\,s$^{-1}$]~~-----------}     
        & \\
\noalign{\smallskip}
\hline\noalign{\smallskip}
   Thin disk   &  35  & 20 & 16 &    0  &  $-15$ &  0.85  \\
   Thick disk  &  67  & 38 & 35 &    0  &  $-46$ &  0.09  \\
   Halo        & 160  & 90 & 90 &    0  & $-220$ &  0.0015 \\
   Hercules    &  26  &  9 & 17 & $-40$ &  $-50$ &  0.06    \\
\hline
\end{tabular}
\flushleft
$^{\dagger}${\scriptsize
		$\sigma_{\rm U}$, 
		$\sigma_{\rm V}$, and $\sigma_{\rm W}$ are the velocity dispersions  for the different components; $U_{\rm LSR}$ and $ V_{\rm LSR}$ are the asymmetric drifts in $U$ and $V$ relative to the LSR; and X is the 
        normalisation fractions for each component in the Solar neighbourhood
        (in the {\respondtoxiang Galactic} plane).
        Values are taken from for the thin disk, thick disk, 
        the stellar halo, and the Hercules stream \citep{2007ApJ...655L..89B,2014A&A...562A..71B}. }
\end{table}

\subsubsection{calibration sample}
\label{sec.meth.rev.cal}
To revise the values of characteristic kinematic parameters, {\xie we rely on a calibration sample based on the LAMOST and Gaia data.}
{\respondtoxiang The LAMOST main-sequence turn-off and subgiant (MSTO-SG) star sample of \citep{2017ApJS..232....2X}} provides the estimates of stellar age, mass, and RV for 0.93 million {\respondtoxiang Galactic}-disk  stars from the LAMOST {\respondtoxiang Galactic} spectroscopic surveys.
The typical uncertainty in age is 34\%. 

To construct the calibration sample, we first cross-matched the above LAMOST MSTO-SG catalog with the Gaia DR2 catalog.
This was done by using the X-match service of CDS. 
We set a critical distance of 1.25 arcseconds for position match.
We carried out a magnitude cut, {\respondtoxiang which was set as} the G magnitude difference less than 2.3,  to ensure our cross-matches were of similar brightness. 
The G magnitudes for the LAMOST MSTO-SG stars were calculated by using the $\rm XSTPS-GAC$ g, r, and i $\rm color-color$ polynomial fits in Table 7 of \cite{2010A&A...523A..48J}. 
For stars with multiple matches, we kept those with the smallest angular separations. 
After the above cross-match, we have 863,663 stars left.

{\xie We then applied the following filters to further clean the calibration sample. } 

  \ {\xie (1) Binary filter. We removed binary star systems because their kinematics contain additional motions \citep{1998MNRAS.298..387D}. This was done by choosing stars flagged as 'Normal star' (i.e., single and with spectral type of AFGKM) in the LAMOST MSTO-SG catalog \citep{2017ApJS..232....2X}. }
  
  \ {\xie (2) Parallax precision filter. Following \cite{1998MNRAS.298..387D}, we removed stars with relative parallax errors larger than 10 percent as reported in the Gaia DR2.}
  
  \ {\xie (3) Age precision filter. We removed stars with age older than 14 Gyr or the error of age larger than 25\% or blue straggler stars  ($|Z|> 1.5$ kpc and age younger than 2 Gyr) in the LAMOST MSTO-SG catalog.} 
  
  \ {\xie (4) Distance filter \citep[similar to][]{1997ESASP.402..473B}.  The majority of the remaining stars are brighter than G mag=16 where the parallax has a median error of 0.0649 mas. Recalling the above 10 percent parallax precision requirement, then it leads to a distance limit  $\sim 1/(0.0649/0.1) = 1.54$ kpc. 
  We therefore removed stars with distance larger than this limit. Such a cut at 1.54 kpc also makes the distance distribution of the calibration sample closer to that of the planet host sample (Figure \ref{figRtheZstars}). }
  
\begin{figure}[!t]
\centering
\includegraphics[width=0.5\textwidth]{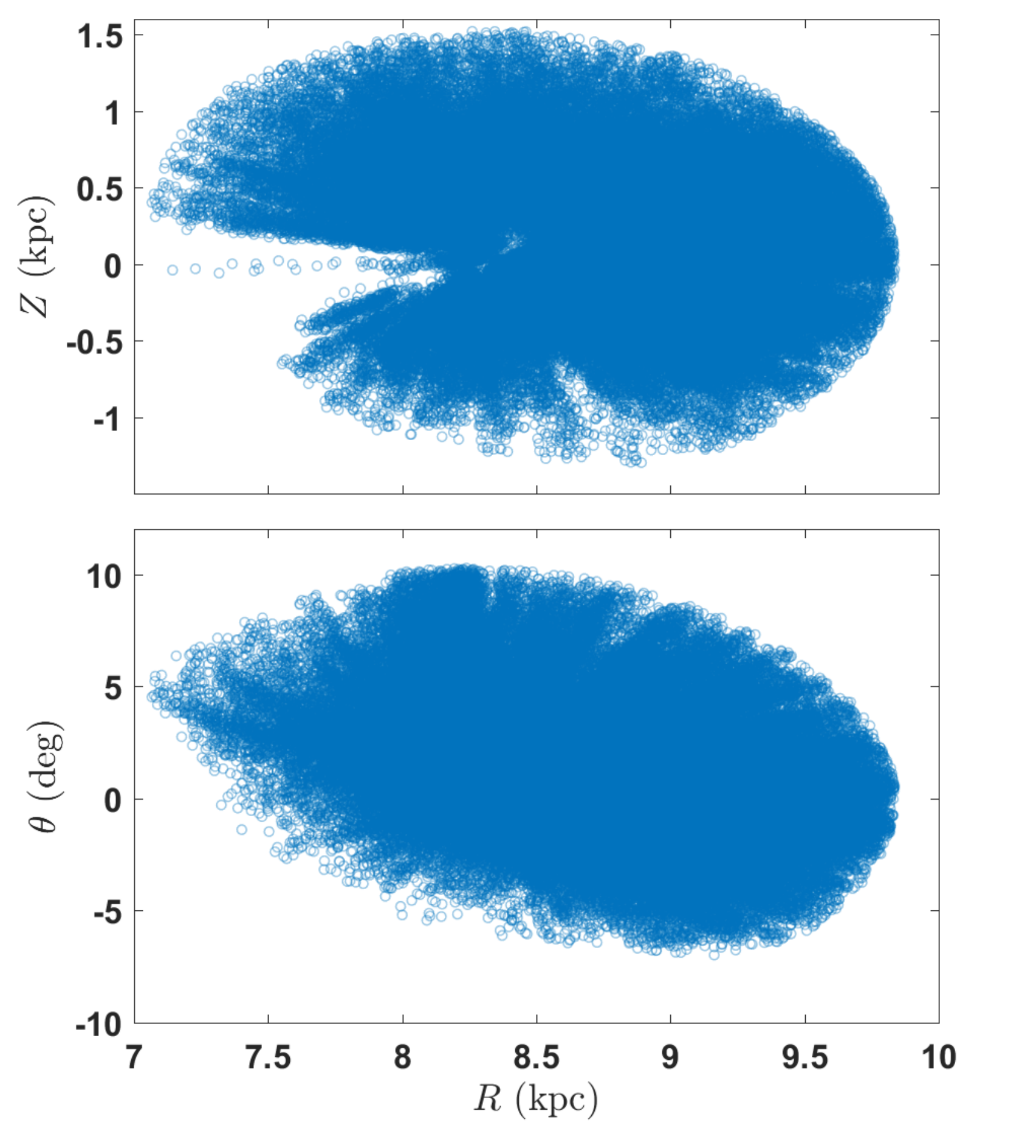}
\caption{Top panel: Galactocentric radius ($R$) vs. height ($Z$) for the calibration sample. (8.34 kpc, 0.027 kpc) marks the position of Sun. 
Bottom panel: Galactocentric radius ($R$) vs. angle ($\theta$) for the calibration sample. (8.34 kpc, 0) marks the position of Sun. 
\label{figMSTORtheZ}}
\end{figure}
  
{\xie After applying the above filters, we are left with 
130,403 stars. Figure \ref{figMSTORtheZ} shows the location of stars in this calibration sample. As can be seen, these stars are mainly located at  $7.5 < R < 10.0$ kpc and $|z|<1.5$ kpc, a region large enough to cover most known planet host stars (Figure \ref{figRtheZstars}). } 

{\jiwei Although most planet hosts are main-sequence stars while stars in the calibration sample are main-sequence turn-off stars and sub-giants, it should not affect the calibration of kinematic properties.
In fact, it has been shown that the velocity ellipsoid \citep{2014MNRAS.439.1231B,2015MNRAS.452..956B,2019MNRAS.489..910E} and the AVR \citep{1977A&A....60..263W,2009A&A...501..941H,2018MNRAS.475.1093Y,2019MNRAS.489..176M} are independent on stellar evolution stage and effective temperature (mass). 
Therefore, the revised characteristic kinematic parameters (section \ref{sec.meth.rev}) and AVR (section \ref{sec.meth.avr}) from the calibration sample can be applicable for stars of different spectral types and evolutionary stages, including the planet host sample (section \ref{sec.res}).

}

\subsubsection{binning and examining the calibration sample}
{\xie In order to calculate the characteristic kinematic parameters for each {\respondtoxiang Galactic} component as a function of ($R$, $Z$) in the Galaxy, we binned the calibration sample as follows.} 
{\xie For $|Z|$, we set 8 bins with boundaries at $|Z|=$ 0, 0.1, 0.2, 0.3, 0.4, 0.55, 0.75, 1.0, and 1.5 kpc, resulting similar sizes ($\sim$ 20,000 stars) for all the bins except for the last two whose sizes are $\sim$ 10,000.}
{\xie For $R$, we set 5 bins with boundaries at $R =$ 7.5, 8.0, 8.5, 9.0, 9.5, and 10 kpc. 
In total, there are $5 \times 8=40$ grids in the $R-Z$ space.
Two of the grids ($R:9.5-10.0$ kpc, $|Z|:0.75-1.0 \ \&\ 1.0-1.5$ kpc) have too few stars ($<400$) and thus are not considered hereafter.}

\begin{figure}[!t]
\centering
\includegraphics[width=0.5\textwidth]{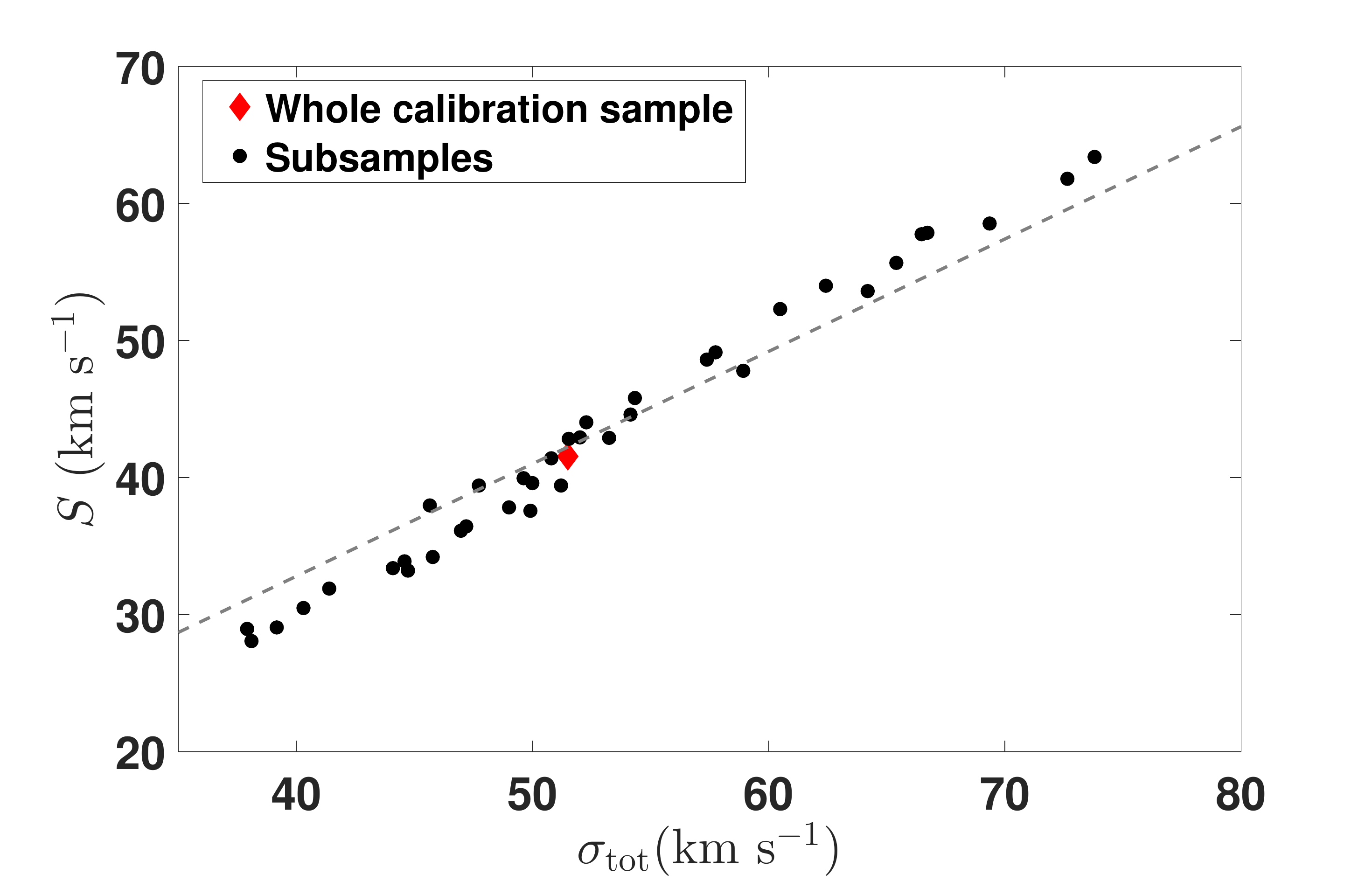}
\caption{The dispersion in proper motions, $S$, of the calibration sample as a
function of the total dispersion in velocity, $\sigma_{\rm tot}$. The black dashed line represents where $S$ and $\sigma_{\rm tot}$ obey Equation \ref{eqSsigmatot}.
\label{figSsigamtot}}
\end{figure}

{\xie Following \cite{2000MNRAS.318..658B},  we examined the kinematically isotropic homogeneity in each bin of the calibration sample.}
{\xie The kinematical homogeneity} requires that the dispersion in proper motions, $S$, follows
       \begin{equation}
       S = (\frac{2}{3}{\sigma_{\rm tot}}^2)^{1/2} = [\frac{2}{3}(\sigma_{R}^2+\sigma_{\theta}^2+\sigma_{Z}^2)]^{1/2}.
       \label{eqSsigmatot}
       \end{equation}
{\xie The result of this examination is shown in Figure \ref{figSsigamtot}.
As can be seen, the calibration sample, either as a whole or as being divided into various grids, generally obeys the above relation, indicating the sample is kinematically unbiased.}

\subsubsection{classifying the calibration sample}
\begin{figure}[!t]
\centering
\includegraphics[width=0.5\textwidth]{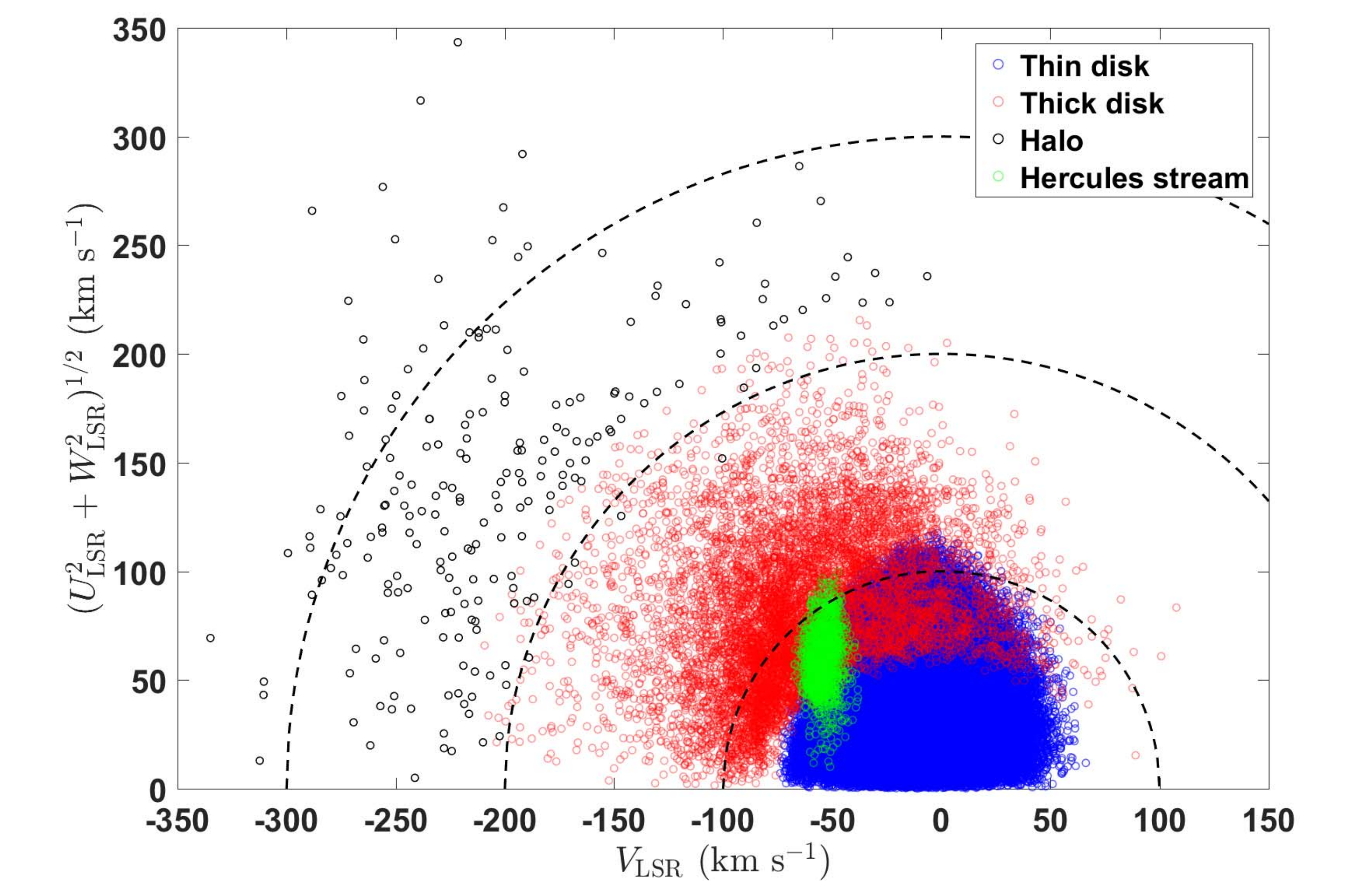}
\caption{The Toomre diagram of the calibration sample for different {\respondtoxiang Galactic} components.
The diagram is colour-coded to represent different components. 
Dashed lines show constant values of the total {\respondtoxiang Galactic} velocity $V_{\rm tot} = 100, \ 200, {\ \rm and} \ 300 \ \rm km \ s^{-1}$.
\label{figMSTOToomre}}
\end{figure}

{\xie Next, we classify stars in the calibration sample into different {\respondtoxiang Galactic} components.}
{\xie Following \citet{2017ApJ...845..101B}, we identify halo stars if $V_{\rm tot} = (U_{\rm LSR}^2+V_{\rm LSR}^2+W_{\rm LSR}^2)^{1/2}>220 \ \rm km \ s^{-1}$}. 
Stars with $V_{\rm LSR} \approx -50 \ \pm \ 9 \ \rm km \ s^{-1}$ and $(U_{\rm LSR}^2+W_{\rm LSR}^2)^{(1/2)} \approx \rm 50-70 \ km \ s^{-1}$ are selected as the Hercules stream  \citep{2005A&A...430..165F,2007ApJ...655L..89B,2014A&A...562A..71B}.
{\xie We adopt the age-defined thin and thick disc components with a boundary at 8 Gyr \citep{1998A&A...338..161F,2013IAUS..292..105H} to classify the rest sample into thin and thick disc stars.
In Figure \ref{figMSTOToomre}, we plot the Toomre diagram for the calibration sample. 
As expected, most stars with low velocities ($V_{\rm tot}\lesssim 50 \rm km\ s^{-1}$) are in thin disk, while those with moderate velocities ($V_{\rm tot}\sim70-180 \ \rm km\ s^{-1}$) are mainly in thick disk (e.g. \cite{2003A&A...397L...1F,2013A&A...554A..44A,2014A&A...562A..71B}. 
}).

\begin{figure*}[!ht]
\centering
\includegraphics[width=\textwidth]{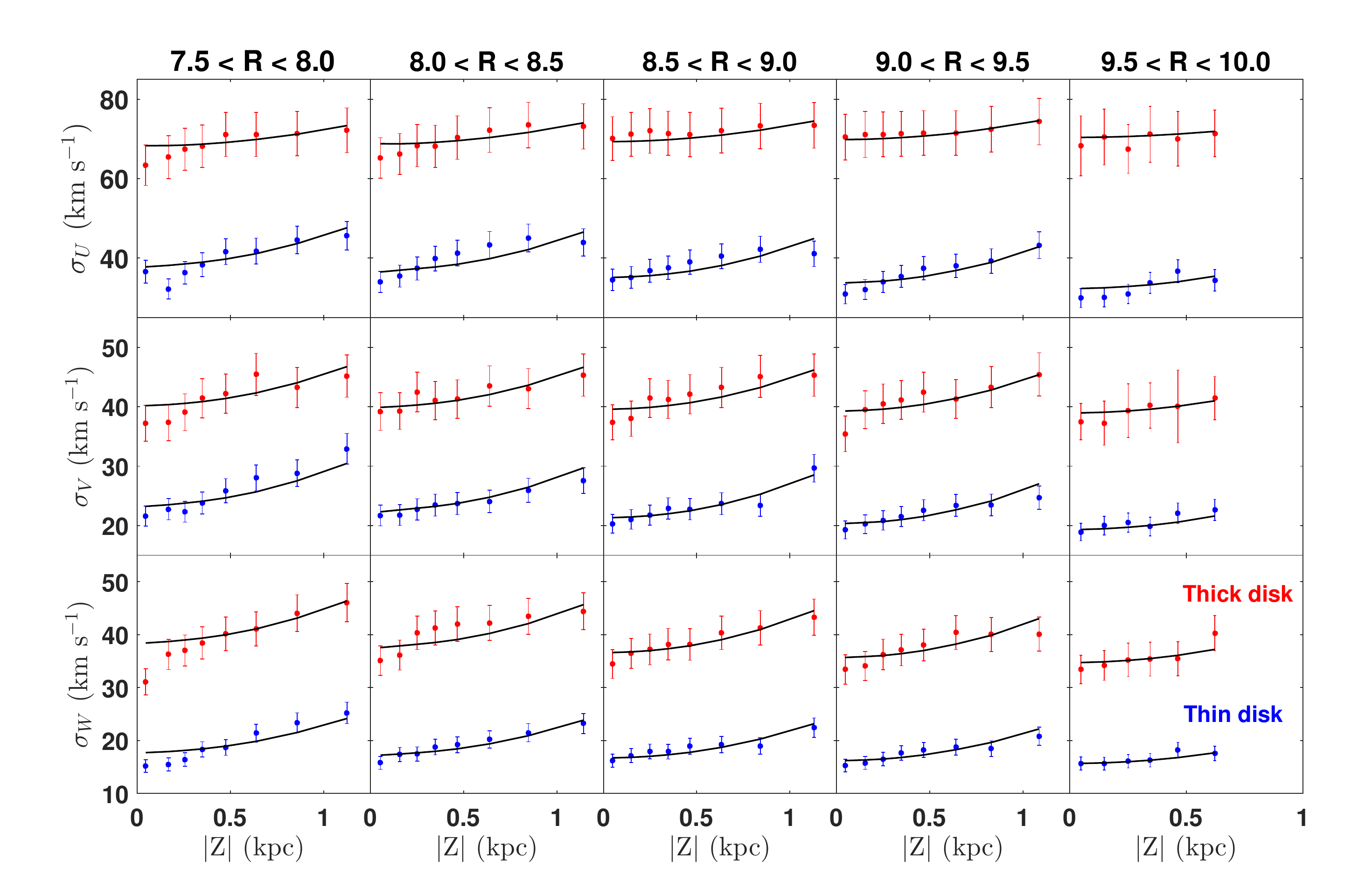}
\caption{The velocity dispersions as functions of position (R, Z) in the Galaxy for the calibration sample. 
The black line in each panel denotes the result of the best fit of Equation \ref{eUVWRZ} using the coefficients in Table \ref{tab:eUVWpara}.
\label{figeUVWRZ}}
\end{figure*}

\subsubsection{revising the velocity ellipsoid}
\begin{figure}[!t]
\centering
\includegraphics[width=0.5\textwidth]{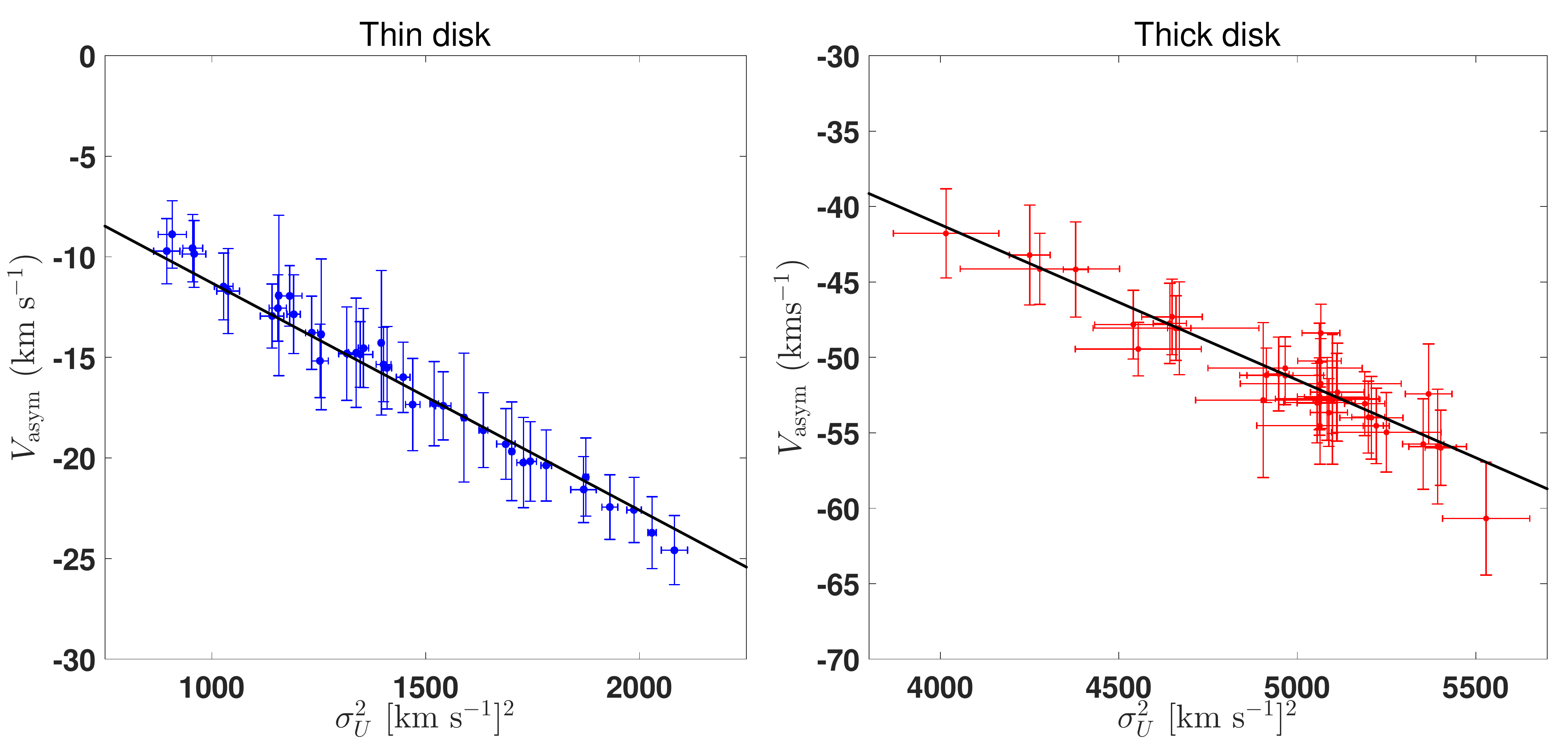}
\caption{The asymmetric velocity, $V_{\rm asym}$ as a function of $\sigma_U^2$ for the thin disk (left panel) and thick disk (right pannel). 
The black lines denote the results of the best fit using Equation \ref{eqVasym}. 
\label{figVasym}}
\end{figure}

{\xie We then revise the velocity ellipsoid of each {\respondtoxiang Galactic} component, namely calculate $\sigma_U$, $\sigma_V$, $\sigma_W$, and $V_{\rm asym}$ of each grid in the $R-Z$ plane.
Here, we revise those values only for the thin and thick disk components. 
For the halo and Hercules components, their numbers in each grid are too low for the revision, thus we adopt the velocity ellipsoid values as in \cite{2007ApJ...655L..89B,2014A&A...562A..71B} instead.
The calculated values of $\sigma_U$, $\sigma_V$, $\sigma_W$, and $V_{\rm asym}$ for the thin and thick disk components are tabulated in Table \ref{tab:dispersionsrevised} and visualized in Figure \ref{figeUVWRZ} and Figure \ref{figVasym}. }

{\xie For velocity dispersion, according to \cite{2013MNRAS.436..101W}, it generally} follows a simple formula:
\begin{equation}
\sigma = b_1 + b_2\times \frac{R}{\rm kpc} + b_3 \times (\frac{Z}{\rm kpc})^2 {\rm km \ s^{-1}}.
\label{eUVWRZ}
\end{equation}
{\xie
We therefore fit $\sigma_U$, $\sigma_V$, $\sigma_W$ in this formula.
{\respondtoWang To obtain the uncertainty of each fitting parameter, we assumed that the {\respondtoxiang Galactic} velocity following the Gaussian distribution $N (V, err\_V)$, where $err$ represents the corresponding uncertainty. 
Then, we resampled {\respondtoxiang Galactic} velocities based on these Gaussian distributions.
After that, we refit the resampled data in the formula of Equation \ref{eUVWRZ}.
We repeated the above resampling process 1,000 times and obtained 1,000 sets of best fits.
The uncertainties (one-sigma interval) of the fitting parameters are set as the range of $50\pm 34.1$ percentiles of these 1,000 sets of best fits. }
Figure \ref{figeUVWRZ} shows the velocity dispersions ($\sigma_U$, $\sigma_V$, $\sigma_W$) as a function of $Z$ in each $R$ bin.
The best fits are over-plotted as the black solid curves.
The values of fitting parameters and their one-sigma uncertainties are summarized in Table \ref{tab:eUVWpara}. 
As expected, velocity dispersions generally increase with $Z$ for both thin and thick disks in all the $R$ bins. 
}

\begin{table}[!t]
\centering
\caption{Fitting parameters of the velocity dispersion as functions of $(R,\  Z)$, {\chen i.e. Equation \ref{eUVWRZ}}.}
{\footnotesize
\label{tab:eUVWpara}
\begin{tabular}{lccc} \hline
        & {$b_1$} & {$b_2$} & {$b_3$}  \\  \hline
   \multirow{2}{*}{$\sigma^{\rm D}_{U}$}  &  \multirow{2}{*}{$63.4^{+1.3}_{-3.2}$} & \multirow{2}{*}{$-3.2^{+0.3}_{-0.2}$} & \multirow{2}{*}{$7.6^{+0.5}_{-1.3}$} \\ \\
   \multirow{2}{*}{$\sigma^{\rm D}_{V}$}  &  \multirow{2}{*}{$41.6^{+0.7}_{-2.3}$} & \multirow{2}{*}{$-2.3^{+0.3}_{-0.1}$} & \multirow{2}{*}{$5.6^{+0.2}_{-0.9}$} \\ \\
   \multirow{2}{*}{$\sigma^{\rm D}_{W}$}  &  \multirow{2}{*}{$27.3^{+1.3}_{-1.5}$} & \multirow{2}{*}{$-1.2^{+0.2}_{-0.1}$} & \multirow{2}{*}{$5.0^{+0.2}_{-0.7}$} \\ \\
   \multirow{2}{*}{$\sigma^{\rm TD}_{U}$}  &  \multirow{2}{*}{$58.4^{+6.7}_{-4.4}$}  & \multirow{2}{*}{$1.2^{+0.6}_{-0.7}$} & \multirow{2}{*}{$4.1^{+0.4}_{-0.8}$} \\ \\
   \multirow{2}{*}{$\sigma^{\rm TD}_{V}$}  &  \multirow{2}{*}{$44.9^{+5.3}_{-2.7}$} & \multirow{2}{*}{$-0.7^{+0.4}_{-0.5}$} & \multirow{2}{*}{$5.2^{+0.3}_{-1.0}$} \\ \\
   \multirow{2}{*}{$\sigma^{\rm TD}_{W}$}  &  \multirow{2}{*}{$55.8^{+1.8}_{-3.1}$} & \multirow{2}{*}{$-2.2^{+0.4}_{-0.2}$} & \multirow{2}{*}{$6.1^{+0.4}_{-1.0}$} \\ \\\hline
   
\end{tabular}}
\end{table}

For asymmetric velocity, {\xie according to \citet{2003A&A...409..523R,2008gady.book.....B},} it generally follows the relation  $V_{\rm asym} = -\frac{\sigma_U^2}{2V_{\rm LSR}}\left[\frac{\partial \ln \rho}{\partial \ln R} +\frac{\partial \ln\sigma_U^2}{\partial \ln R} + \left( 1-\frac{\sigma_V^2}{\sigma_U^2}\right) + \left( 1-\frac{\sigma_W^2}{\sigma_U^2}\right) \right]$.
Therefore, we use the following formula to calculate $V_{\rm asym}$, i.e.,
\begin{equation}
V_{\rm asym} = \sigma_U^2/C_0.
\label{eqVasym}
\end{equation}
Figure \ref{figVasym} shows $V_{\rm asym}$ as a function of $\sigma_U^2$.
{\xie The best fits are over-plotted as the black solid lines.
The values of $C_0$ are $-88.5^{+1.7}_{-1.9} \ \rm km \ s^{-1}$ and $-92.5^{+2.3}_{-2.1} \ \rm km \ s^{-1}$ for the thin and thick disks respectively, which are generally consistent with the theoretical estimate ($-82\pm 6 \ \rm km \ s^{-1}$) \citep{2008gady.book.....B}.
}

\subsubsection{revising the $X$ factor}

\begin{figure*}[!t]
\centering
\includegraphics[width=\textwidth]{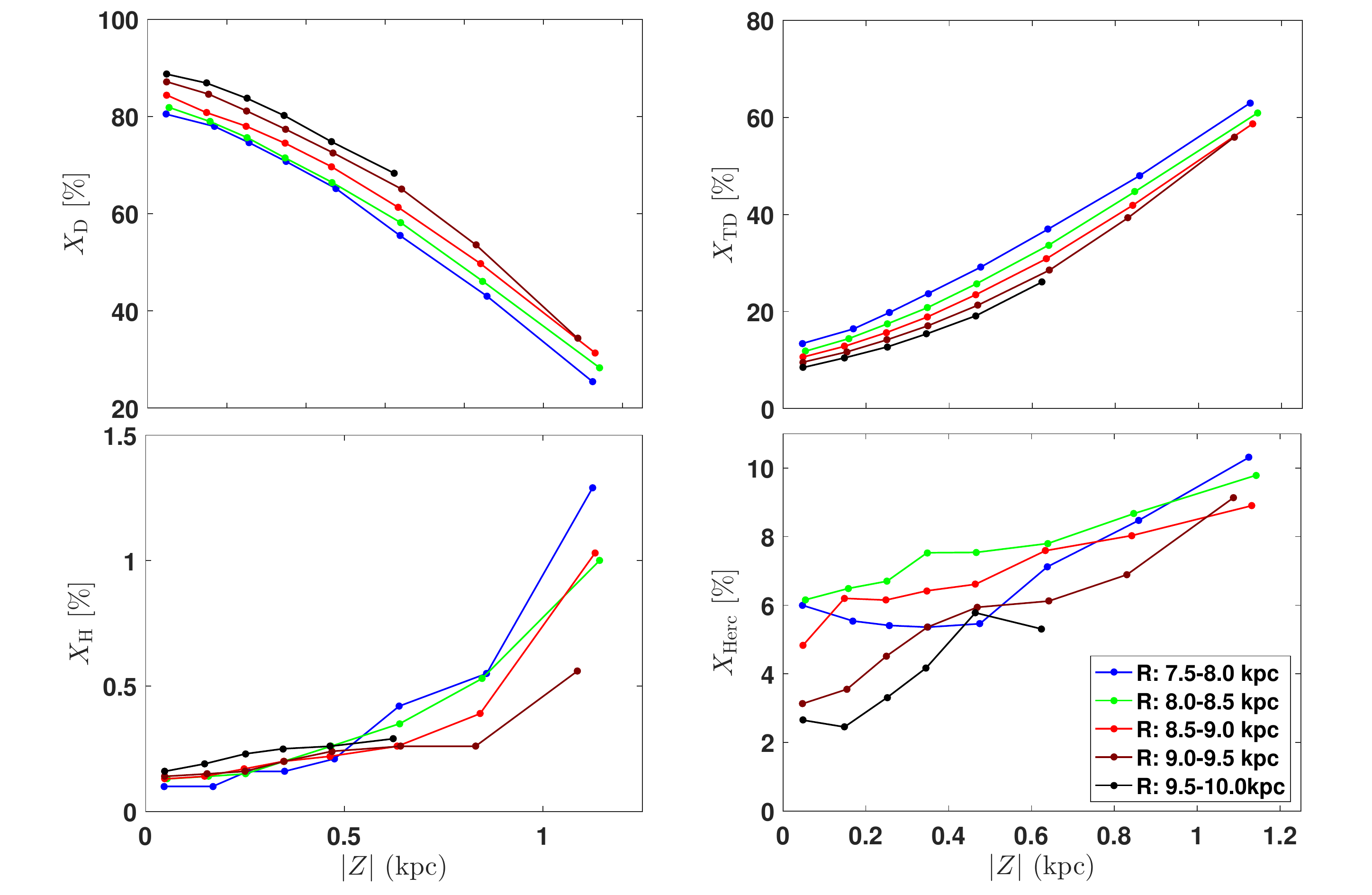}
\caption{The normalisation fraction $X$ of stars for each component as a function of {\respondtoxiang Galactic} radius $R$ and absolute value of height $|Z|$. The different colours denote subsamples of stars with different Galactic radii.
\label{fXDTDHHer}}
\end{figure*}

{\xie As defined in Equation \ref{eqTDD},  $X$ is the fraction of stars for a given component. 
For halo and Hercules stream, their number density distributions and structures are not quite clear yet. 
Thus, we set their $X$ as the observed fractions, i.e.,
\begin{equation}
 X_{\rm H} =\frac{N_{\rm H}}{N_{\rm tot}},  \  X_{\rm Herc} =\frac{N_{\rm Herc}}{N_{\rm tot}},
\label{XHHer}
\end{equation}
where $N_{\rm H}$, $N_{\rm Herc}$ and $N_{\rm tot}$ are the numbers of Hercules stream stars, halo stars, and total stars in each $(R-Z)$ grid. 
}

For thin and thick disks, {\xie the number density is modeled as the following formula}  \citep{2001ApJ...553..184C,2008gady.book.....B}: 
\begin{equation}
n(R,Z) = n^0 \times {\rm exp}(-\frac{R-R_{\odot}}{h_R}){\rm exp}(-\frac{|Z|}{h_{Z}}),
\label{nRZ}
\end{equation}
where $h_Z$ and $h_R$ are the scale height and scale length of the disk, respectively. 
Here we take ($h_R$, $h_z$) as (3.4, 0.3) kpc for the thin disk and (1.8, 1.0) kpc for the thick disk \citep{2008gady.book.....B,2012ApJ...752...51C,2016ApJ...823...30B}. 
{\xie Then, the ratio of thick/thin disk star numbers in each $R-Z$ grid can be calculated as}
\begin{equation}
X_{\rm TD/D}=X_{\rm TD}/X_{\rm D} = \frac{\int_{R} \int_{Z} {n_{\rm TD}(R,Z)} 2\pi R{\rm d}R{\rm d}Z}{\int_{R} \int_{Z} {n_{\rm D}(R,Z)} 2\pi R{\rm d}R{\rm d}Z}.  
\label{eqNDTD}
\end{equation}
 
In practice, we first calculated nominal ratio, i.e., ${n^0}_{\rm TD}/{n^0}_{\rm D}$=0.098 (first term of the right-hand side of Equation \ref{nRZ}) by solving Equation \ref{eqNDTD} with $X_{\rm TD}/X_{\rm D} = 0.09/0.85$ for the solar neighbourhood ($((R-R_\odot)^2+Z^2)^{1/2}=100$ pc, \cite{2014A&A...562A..71B}).
Then, we applied this nominal ratio to Equation \ref{eqNDTD} to calculate the ratio of thick/thin disk star numbers in each $R-Z$ grid. 

Finally, the $X$ of thin and thick disks were calculated as:
\begin{equation}
\begin{aligned}
 &X_{\rm D} = (1-X_{\rm H}-X_{\rm Herc}) \times \frac{1}{1+X_{\rm TD/D}},  \\
 &X_{\rm TD} = (1-X_{\rm H}-X_{\rm Herc}) \times \frac{X_{TD/D}}{1+X_{\rm TD/D}}.
\label{XDTD}
\end{aligned}
\end{equation}

{\xie The results of these revised $X$ values in all the $R-Z$ grids are tabulated in Table \ref{tab:dispersionsrevised}.}
Figure \ref{fXDTDHHer} shows the X values of various {\respondtoxiang Galactic} components as functions of {\respondtoxiang Galactic} radius $R$ and absolute value of height, $|Z|$. 
As expected, $X_{\rm D}$ ($X_{\rm TD}$) generally decrease (increase) with $|Z|$ in all the $R$ bins.

\subsubsection{comparison to \cite{2014A&A...562A..71B} }
{\xie
So far, we have calculated the characteristic parameters (i.e., $\sigma_U$, $\sigma_V$, $\sigma_W$, $U_{\rm asym}$, and $V_{\rm asym}$) as functions of $R$ and $Z$ (Table \ref{tab:dispersionsrevised}).
In Table \ref{tab:KPcomparison}, we then compare our results in the solar neighbourhood (here bin with $R:8.0-8.5$ kpc and  $|Z|<0.1$ kpc) to those of \cite{2014A&A...562A..71B}.
As can be seen, both our results and those of \citet{2014A&A...562A..71B} provide very similar values on these characteristic parameters, demonstrating that our revision (section \ref{sec.meth.classify}) can also be reduced to the solar neighbourhood. 
}

\begin{table}[!t]
\centering
\caption{The kinematic characteristics for stellar components in the Solar neighbourhood from \cite{2014A&A...562A..71B} and this work.}
{\footnotesize
\label{tab:KPcomparison}
\begin{tabular}{l|cccc} \hline
  &  $X_{\rm D}$   & $X_{\rm TD}$ & $X_{\rm H}$ & $X_{\rm Herc}$ \\ \hline
Bensby et al. (2014)  & 0.85 & 0.09 & 0.0015 & 0.06 \\
This work & 0.84 & 0.10 & 0.0013 & 0.06 \\ \hline
&  $\sigma_{\rm U\_D}$   & $\sigma_{\rm V\_D}$ & $\sigma_{\rm W\_D}$ & $V_{\rm asymD}$ \\ \hline
Bensby et al. (2014) & 35 & 20 & 16 & -15 \\
This work & 34 & 21 & 16 & -14 \\ \hline
&  $\sigma_{\rm U\_TD}$   & $\sigma_{\rm V\_TD}$ &  $\sigma_{\rm W\_TD}$ & $V_{\rm asymTD}$ \\ \hline
Bensby et al. (2014) & 67 & 38 & 35 & -46 \\
This work & 65 & 39 & 35 & -44 \\ \hline

\end{tabular}}
\end{table}

\section{\chen Revisiting the Age-Velocity Dispersion Relation to Derive Kinematic Ages}
\label{sec.meth.avr}
When the stars in the solar neighbourhood are binned by age, the velocity dispersion of each bin increases with its age. 
This age velocity relation (AVR) has been known and studied for decades \citep{1946ApJ...104...12S,1950AZh....27...41P,1977A&A....60..263W,2009A&A...501..941H}. 
Similar relationship has also been inferred for external {\respondtoxiang Galactic} disk  \citep{2016MNRAS.462.1697A,2017A&A...605A...1R}. 
{\xie Here, we revisit the AVR with the calibration sample constructed in section \ref{sec.meth.rev.cal}.}


\subsection{Fitting AVR}
\label{sec.meth.AVR.fit}
\begin{figure*}[!ht]
\centering
\includegraphics[width=\textwidth]{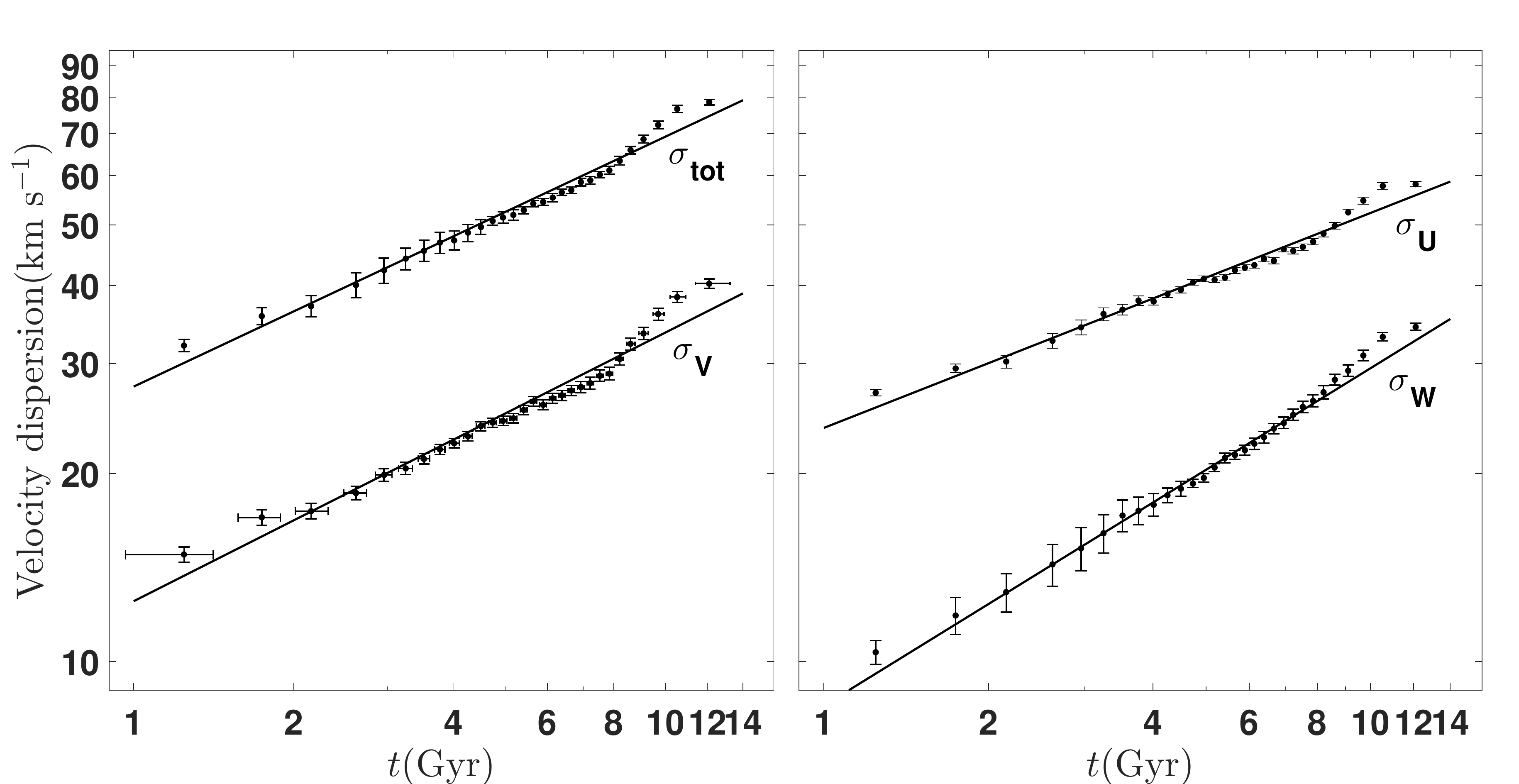}
\caption{The Velocity dispersions for $U_{\rm LSR}, V_{\rm LSR}, W_{\rm LSR}$ and $V_{\rm tot}$ vs. age for the selected calibration star sample.
The 30 bins have approximately equal numbers of stars ($\sim 4350$ in each); 
The solid black lines denote the respective best fit of refitting AVR (Equation \ref{AVR}) using the coefficients in Table \ref{tab:AVRpara}.
\label{fAVR}}
\end{figure*}

In our study, {\xie we} divided the foregoing calibration sample (section \ref{sec.meth.rev.cal}) into 30 bins with approximately equal sizes ($\sim$ 4,350 stars in each bins) according to their ages. 
{\xie Then we calculated the total velocity dispersion for each bin, i.e.,
\begin{equation}
\sigma_{\rm tot} =  (\sigma_U^2+\sigma_V^2+\sigma_W^2)^{\frac{1}{2}}.
\label{sigmatot}
\end{equation}
Figure \ref{fAVR} shows the velocity dispersion as a function of the median age of each bin.}
As can be seen, {\xie all the components of velocity dispersion ($U_{\rm LSR}, V_{\rm LSR}, W_{\rm LSR}$) and the total velocity dispersion ($V_{\rm tot}$) increase with age.}
{\xie Following \citet{2009A&A...501..941H,2016MNRAS.462.1697A}, we fit the AVRs shown in Figure \ref{fAVR} by using a simple power law formula, i.e.,}
\begin{equation}
\sigma =  k \times \left(\frac{t}{\rm Gyr} \right)^{\beta} \, \rm km \ s^{-1},
\label{AVR}
\end{equation}
where t is stellar age, $\sigma$ is the velocity dispersion,  $k$ and $\beta$ are two fitting parameters.

{\xie We used the Levenberg-Marquardt algorithm (LMA) to find the best fit.} 
To obtain the uncertainty of each fitting parameter, we assumed that the {\respondtoxiang Galactic} velocity and age follows the Gaussian distribution $N (V, err\_V)$ and $N (t, err\_t)$, where $err$ represents the corresponding uncertainty. 
Then, we resampled {\respondtoxiang Galactic} velocities and stellar ages based on these Gaussian distributions.
After that, we refit the AVR by using the resampled data.
We repeated the above resampling process 1,000 times and obtained 1,000 sets of best fits.
The uncertainties (one-sigma interval) of the fitting parameters are set as the range of $50\pm 34.1$ percentiles of these 1000 sets of best fits. 

The values of the fitting parameters on the AVR are summarized in Table \ref{tab:AVRpara}. 
{\xie We obtained $\beta$=$0.34^{+0.02}_{-0.01}$, $0.43^{+0.02}_{-0.02}$, $0.54^{+0.02}_{-0.02}$, and $0.40^{+0.02}_{-0.02}$, for $U_{\rm LSR}, V_{\rm LSR}, W_{\rm LSR}$, and $V_{\rm tot}$, respectively, which are consistent with the values derived from previous studies \citep[][see section \ref{sec.meth.AVR.com} for detail comparisons]{2009A&A...501..941H, 2009MNRAS.397.1286A}.}

\begin{table*}[!ht]
\centering
\caption{Fitting parameters of the Age-Velocity dispersion Relation (AVR, Equation \ref{AVR}).}
{\footnotesize
\label{tab:AVRpara}
\begin{tabular}{lcccccccc} \hline
        & \multicolumn{2}{c}{---------~~$U$~~---------} & \multicolumn{2}{c}{---------~~$V$~~---------} & \multicolumn{2}{c}{---------~~$W$~~---------} & \multicolumn{2}{c}{---------~~$V_{\rm tot}$~~---------}   \\ 
        &  $k$  & $\beta$ &  $k$  & $\beta$ &  $k$  & $\beta$ &  $k$  & $\beta$ \\ \hline \hline
\multicolumn{9}{c}{section \ref{sec.meth.AVR.fit}: AVR of the whole calibration sample} \\ \hline
\multirow{2}{*}{Whole sample} & \multirow{2}{*}{$23.66^{+0.66}_{-0.59}$} & \multirow{2}{*}{$0.34^{+0.02}_{-0.01}$} & \multirow{2}{*}{$12.49^{+0.49}_{-0.44}$} & \multirow{2}{*}{$0.43^{+0.02}_{-0.02}$} & 
\multirow{2}{*}{$8.50^{+0.47}_{-0.41}$} & 
\multirow{2}{*}{$0.54^{+0.02}_{-0.02}$} & \multirow{2}{*}{$27.55^{+0.82}_{-0.71}$} & \multirow{2}{*}{$0.40^{+0.02}_{-0.02}$} \\ \\ \hline \hline
\multicolumn{9}{c}{section \ref{sec.meth.AVR.rad}: AVR of different radii} \\ \hline
   \multirow{2}{*}{$R: 7.5-8.0$ kpc} &  \multirow{2}{*}{$24.16^{+0.17}_{-0.12}$} & \multirow{2}{*}{$0.33^{+0.01}_{-0.01}$} & \multirow{2}{*}{$14.32^{+0.10}_{-0.10}$} &  \multirow{2}{*}{$0.39^{+0.01}_{-0.01}$} & \multirow{2}{*}{$8.74^{+0.08}_{-0.07}$} & \multirow{2}{*}{$0.57^{+0.01}_{-0.01}$} & \multirow{2}{*}{$30.90^{+0.16}_{-0.12}$} & \multirow{2}{*}{$0.36^{+0.01}_{-0.01}$} \\ \\ 
   \multirow{2}{*}{$R: 8.0-8.5$ kpc} &  \multirow{2}{*}{$25.31^{+0.15}_{-0.09}$} & \multirow{2}{*}{$0.33^{+0.01}_{-0.01}$} & \multirow{2}{*}{$14.18^{+0.06}_{-0.06}$} & \multirow{2}{*}{$0.39^{+0.01}_{-0.01}$} & \multirow{2}{*}{$8.84^{+0.07}_{-0.06}$} & \multirow{2}{*}{$0.54^{+0.02}_{-0.02}$} & 
   \multirow{2}{*}{$29.91^{+0.07}_{-0.10}$} & \multirow{2}{*}{$0.38^{+0.01}_{-0.01}$} \\  \\
   \multirow{2}{*}{$R: 8.5-9.0$ kpc}  &  \multirow{2}{*}{$23.40^{+0.13}_{-0.15}$} & \multirow{2}{*}{$0.34^{+0.01}_{-0.01}$} & \multirow{2}{*}{$11.84^{+0.03}_{-0.03}$} &  \multirow{2}{*}{$0.46^{+0.01}_{-0.01}$} & \multirow{2}{*}{$8.30^{+0.05}_{-0.04}$} & \multirow{2}{*}{$0.55^{+0.01}_{-0.01}$} & 
   \multirow{2}{*}{$27.05^{+0.20}_{-0.11}$} & \multirow{2}{*}{$0.40^{+0.005}_{-0.003}$} \\ \\ 
   \multirow{2}{*}{$R: 9.0-9.5$ kpc} & \multirow{2}{*}{$22.28^{+0.09}_{-0.10}$}
   & \multirow{2}{*}{$0.35^{+0.02}_{-0.01}$} & \multirow{2}{*}{$12.27^{+0.04}_{-0.05}$} & \multirow{2}{*}{$0.42^{+0.01}_{-0.01}$} &
   \multirow{2}{*}{$8.32^{+0.05}_{-0.05}$} & \multirow{2}{*}{$0.54^{+0.02}_{-0.02}$} & 
   \multirow{2}{*}{$26.42^{+0.11}_{-0.10}$} & \multirow{2}{*}{$0.40^{+0.02}_{-0.01}$} \\ \\
   \multirow{2}{*}{$R:9.5-10.0$ kpc} & \multirow{2}{*}{$22.73^{+0.22}_{-0.25}$}
   & \multirow{2}{*}{$0.32^{+0.03}_{-0.03}$} & \multirow{2}{*}{$12.66^{+0.15}_{-0.15}$} &  \multirow{2}{*}{$0.44^{+0.04}_{-0.04}$} & \multirow{2}{*}{$8.27^{+0.08}_{-0.08}$} & \multirow{2}{*}{$0.53^{+0.04}_{-0.04}$} & 
   \multirow{2}{*}{$27.74^{+0.18}_{-0.21}$} & \multirow{2}{*}{$0.38^{+0.02}_{-0.02}$} \\ \\ \hline  \hline
 \multicolumn{9}{c}{section \ref{sec.meth.AVR.ver}: AVR of $|Z|<0.1$ kpc} \\ \hline 
 \multirow{2}{*}{$|Z|<0.1$ kpc} & \multirow{2}{*}{$23.15^{+0.61}_{-0.45}$}
 & \multirow{2}{*}{$0.33^{+0.01}_{-0.01}$} & \multirow{2}{*}{$12.43^{+0.41}_{-0.40}$} & \multirow{2}{*}{$0.42^{+0.01}_{-0.02}$} & \multirow{2}{*}{$7.98^{+0.40}_{-0.36}$} &
 \multirow{2}{*}{$0.56^{+0.02}_{-0.02}$} & \multirow{2}{*}{$27.01^{+0.68}_{-0.59}$} & \multirow{2}{*}{$0.39^{+0.02}_{-0.01}$} \\ \\ \hline \hline
  \multicolumn{9}{c}{AVR of Holmberg et al. (2009)$^1$} \\ \hline 
 \multirow{2}{*}{Holmberg et al. (2009)} & \multirow{2}{*}{$22.36^{+2.45}_{-2.08}$}
 & \multirow{2}{*}{$0.39^{+0.07}_{-0.09}$} & \multirow{2}{*}{$12.14^{+1.25}_{-1.04}$} & \multirow{2}{*}{$0.40^{+0.06}_{-0.05}$} & \multirow{2}{*}{$8.35^{+1.22}_{-1.03}$} &
 \multirow{2}{*}{$0.53^{+0.08}_{-0.05}$} & \multirow{2}{*}{$25.62^{+2.80}_{-2.68}$} & \multirow{2}{*}{$0.40^{+0.07}_{-0.05}$} \\ \\\hline

   
\end{tabular}}
\flushleft
{\scriptsize
Note 1: Here the parameters and their uncertainties were obtained by fitting the data of \cite{2009A&A...501..941H} with our methods shown in section \ref{sec.meth.AVR.fit}. 
\\}
\end{table*}

\subsection{AVRs of Different {\respondtoxiang Galactic} Components}
\label{sec.meth.AVR.diffcom}
Our calibration sample consists of 98,486 (75.51\%) thin disk stars 23,572 (18.08\%) thick disk stars, 8,152 (6.25\%) Hercules stars, and 193 (0.15\%) halo stars. 
To explore the difference of AVRs between different {\respondtoxiang Galactic} components, for each component, we divided stars into bins with approximately equal size according to their ages.
Due to sample size, the bin numbers are set as 20, 10, 10, and 5 for thin disk, thick disk, Hercules stream, and halo respectively.
For each component, we performed the same method as in section \ref{sec.meth.AVR.fit} to obtain the AVR.

\begin{figure*}[!t]
\centering
\includegraphics[width=\textwidth]{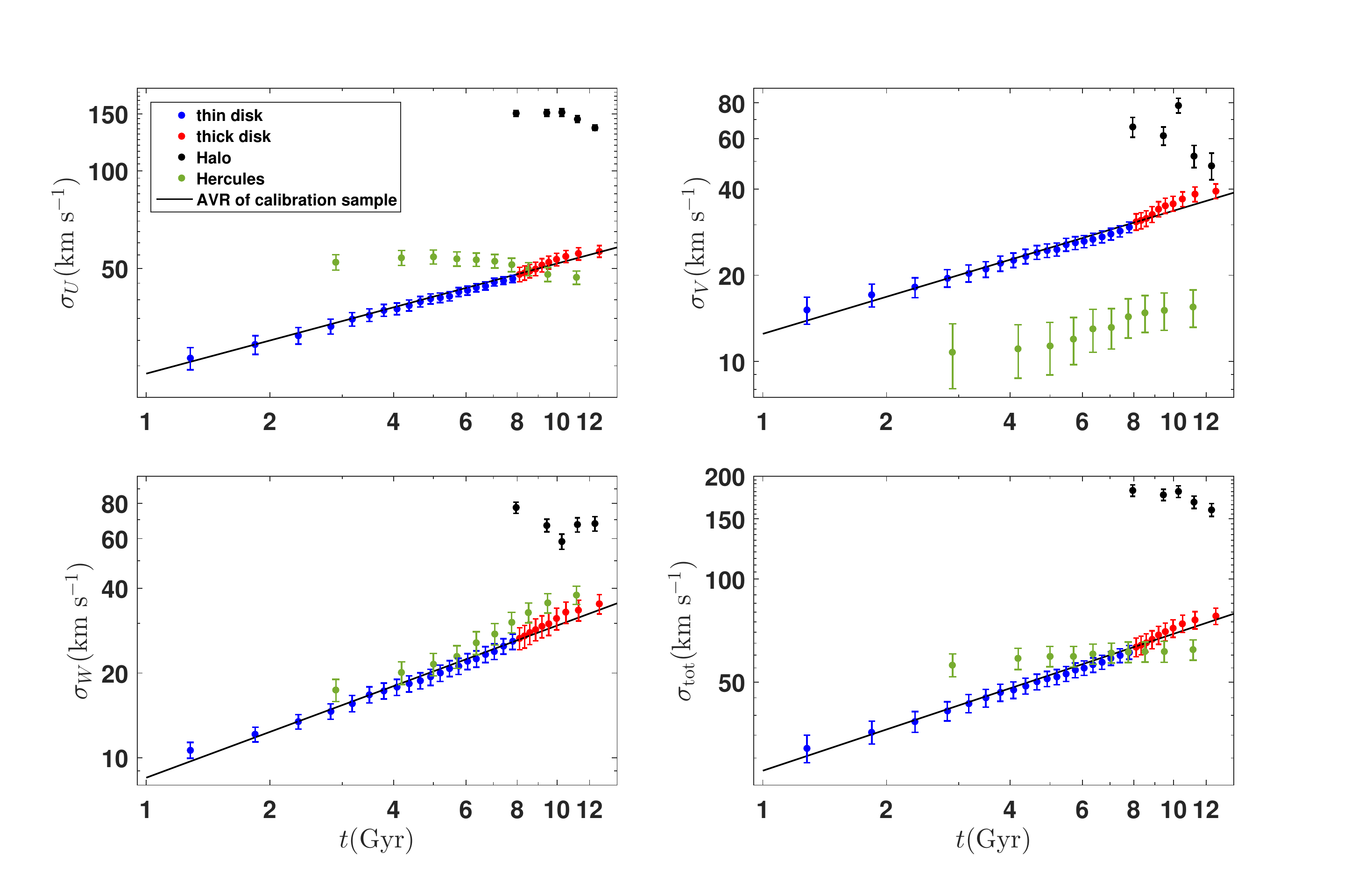}
\caption{The AVRs of differnt {\respondtoxiang Galactic} components. The different colours denote subsamples of stars with different components. The black lines represents the best fit of AVR obtained from the whole sample.
\label{figAVRconponents}}
\end{figure*}

Figure \ref{figAVRconponents} displays the AVRs of different components. 
As can be seen, the AVRs obtained from the thin and thick disk components {\xie fit well to the power-law AVR derived from the whole calibration sample.
For the Hercules stream, the dispersion of velocity component $\sigma_W$ generally follows the  power-law AVR, but other components, i.e, $\sigma_U$ and $\sigma_V$ seem to be irrelevant with age.
For the halo, the velocity dispersions are much larger than those predicted by the power-law AVR, and there is no clear trend between velocity dispersions and ages.
Therefore, we conclude that the power-law AVR can only apply to the thin/thick disk components.}

\subsection{Radial Variation of AVR}
\label{sec.meth.AVR.rad}
\begin{figure}[!h]
\centering
\includegraphics[width=0.5\textwidth]{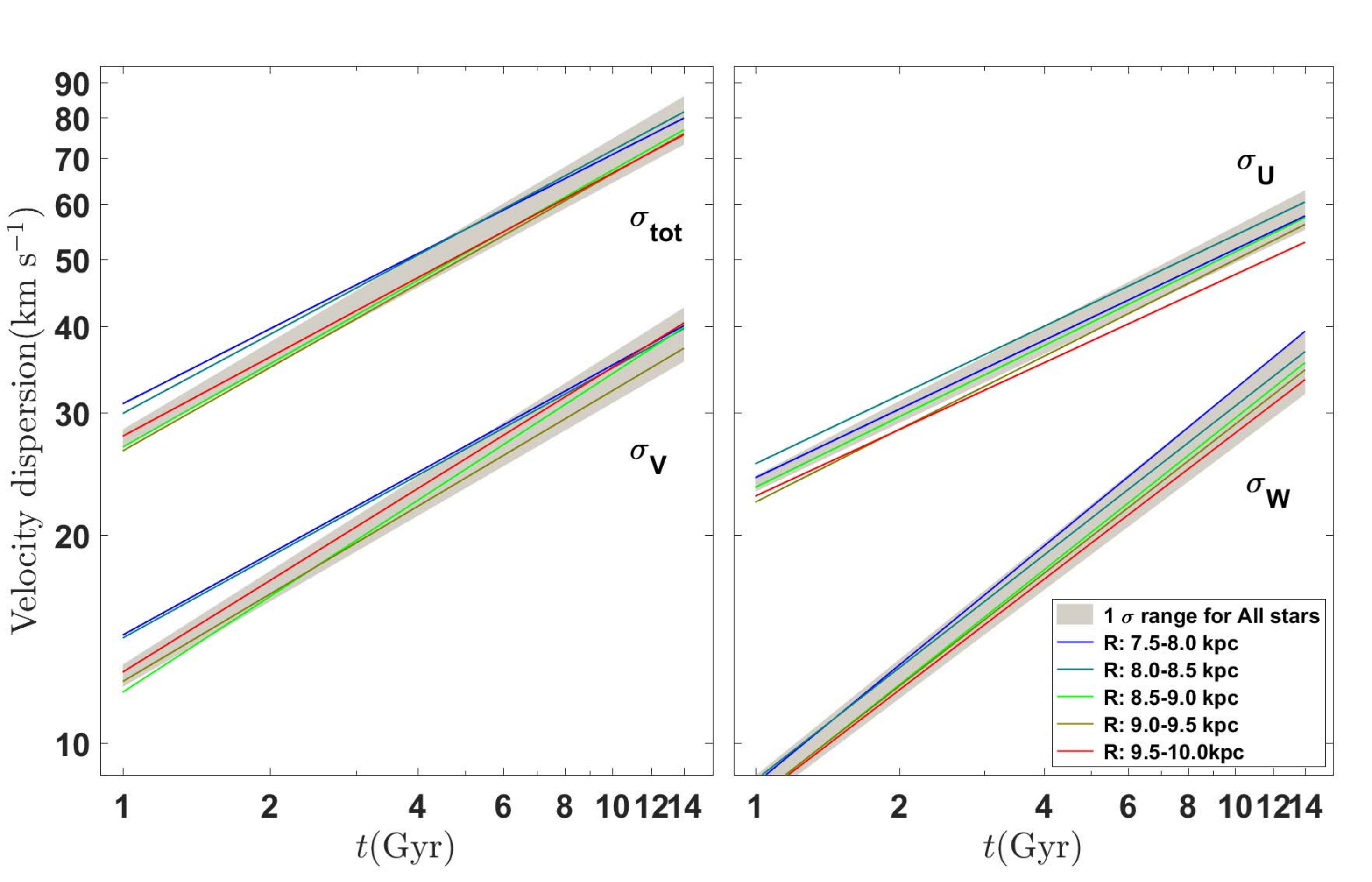}
\caption{The radial variation of AVRs. The different colours denote subsamples of stars with different Galactic radii. The grey region represents the $1-\sigma$ range of AVR obtained from the whole sample.
\label{fAVR_R}}
\end{figure}

{\xie To explore how the AVR varies with {\respondtoxiang Galactic} radius, we divided the foregoing calibration sample (section \ref{sec.meth.rev.cal}) into to five subsamples according to their {\respondtoxiang Galactic} radii, i.e.,  $R=7.5-8.0$ kpc,  $R=8.0-8.5$ kpc, $R=8.5-9.0$ kpc, $R=9.0-9.5$ kpc, and $R=9.5-10$ kpc.}
{\xie For each subsample, we performed the same method as in section \ref{sec.meth.AVR.fit} to fit the AVR.}
Figure \ref{fAVR_R} shows the fitted AVRs for the five subsamples. 
The fitting parameters {\xie of the five {\respondtoxiang AVRs} are} summarized in Table \ref{tab:AVRpara}. 
As can be seen, the AVRs are generally lower with the increase of $R$, which is caused by the decrease of velocity dispersion with $R$ (Equation \ref{eUVWRZ}).
Therefore, the typical uncertainties in $k$, $\beta$ of AVRs 
obtained from the whole sample are generally larger than the five subsamples due to the radial variation.
However, the values of $k$ and $\beta$ differ mildly with R (mean value: $\sim$ 5\% for $k$ and 4\% for $\beta$). 
As can be seen in Figure \ref{fAVR_R}, the AVRs for the five subsamples are generally within $1-\sigma$ range of that for the whole sample (grey regions).
This is consistent with the result derived from simulations in \cite{2016MNRAS.462.1697A}, {\xie which showed that} the shape of the AVR is almost independent of $R$ when $R \gtrsim 6 \rm kpc$. 

\subsection{Vertical Selection Effect of AVR}
\label{sec.meth.AVR.ver}
\begin{figure}[!t]
\centering
\includegraphics[width=0.5\textwidth]{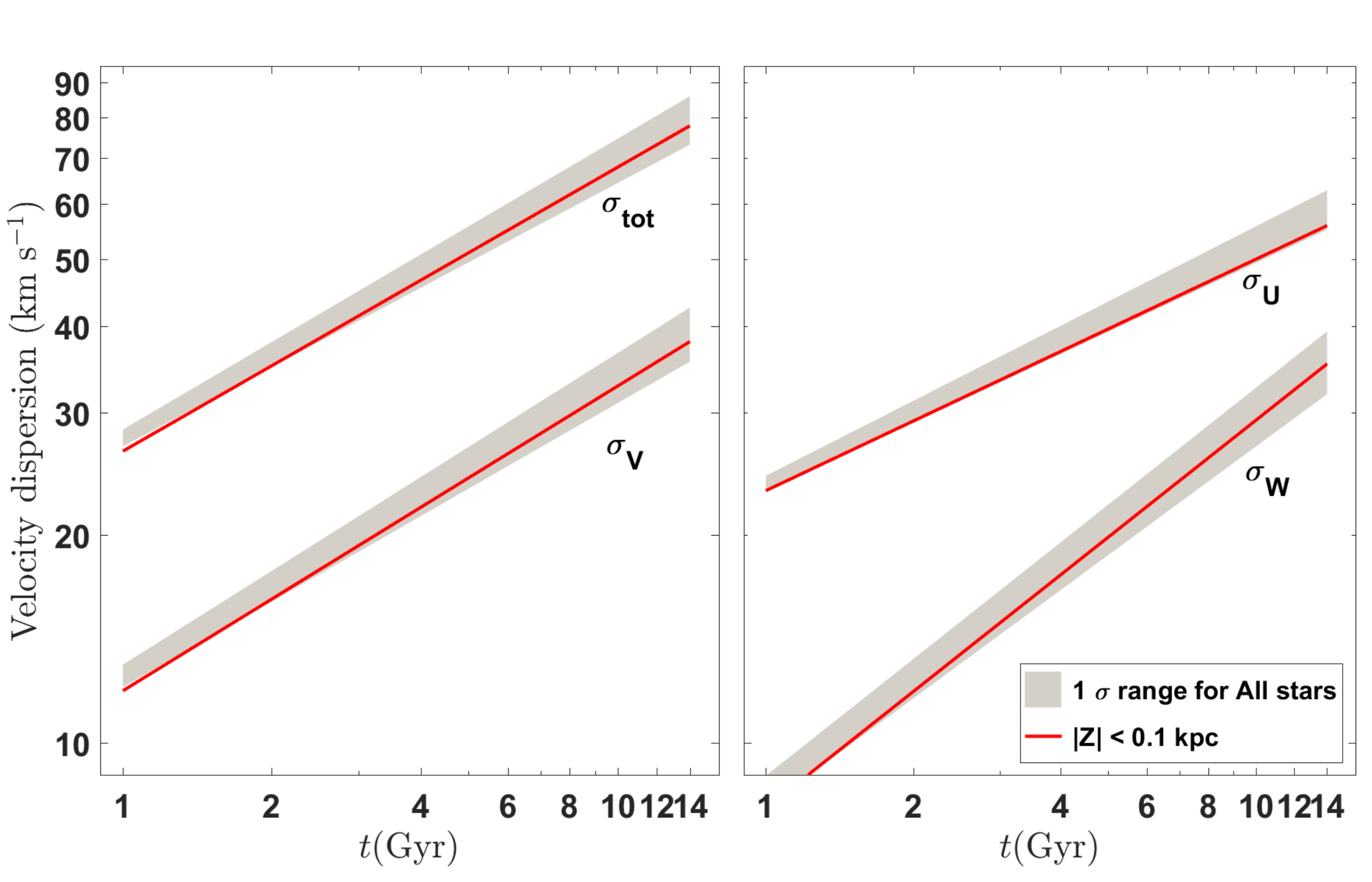}
\caption{Effect of vertical selection on velocity dispersions. The velocity dispersions are plotted as the solid red line for $|z| < 100$ pc. 
The grey region represents the $1-\sigma$ range of AVR obtained from the whole sample.
\label{fAVRZ}}
\end{figure}

Selecting stars from a limited vertical volume introduces phase correlations between stars which influence the values of velocity dispersions at the time of selection and before \citep{2016MNRAS.462.1697A}. 
For example, when stars
are selected close to $Z = 0$, they are all close to their maxima in $|W_{\rm LSR}|$. 
Tracking them back in time, their vertical velocity dispersions thus have to be lower than at the time of selection \citep{2015MNRAS.454.3166A}. 
To explore the vertical selection effect in our sample, {\xie following the} method in \cite{2016MNRAS.462.1697A}, we compared the AVRs for stars with $|Z| < 100$ pc and for all stars irrespective of the $Z$ position. 
The result is displayed in Figure \ref{fAVRZ}. 
As can be seen, there is no significant difference between the AVR for $|Z| < 100$ pc (red) and {\xie that} for all stars (grey region). 
We conducted KS test between the velocity dispersions of stars with $|Z| < 100$ pc and {\xie those of } all stars. 
The p-values are 0.86, 0.97, 0.99, and 0.94 for $U_{\rm LSR}, V_{\rm LSR}, W_{\rm LSR}$, and $V_{\rm tot}$, respectively. 
Such high p-value values demonstrate the vertical selection effect has little influence in AVR.

\subsection{Comparison to Previous Works}
\label{sec.meth.AVR.com}
\begin{figure}[!t]
\centering
\includegraphics[width=0.5\textwidth]{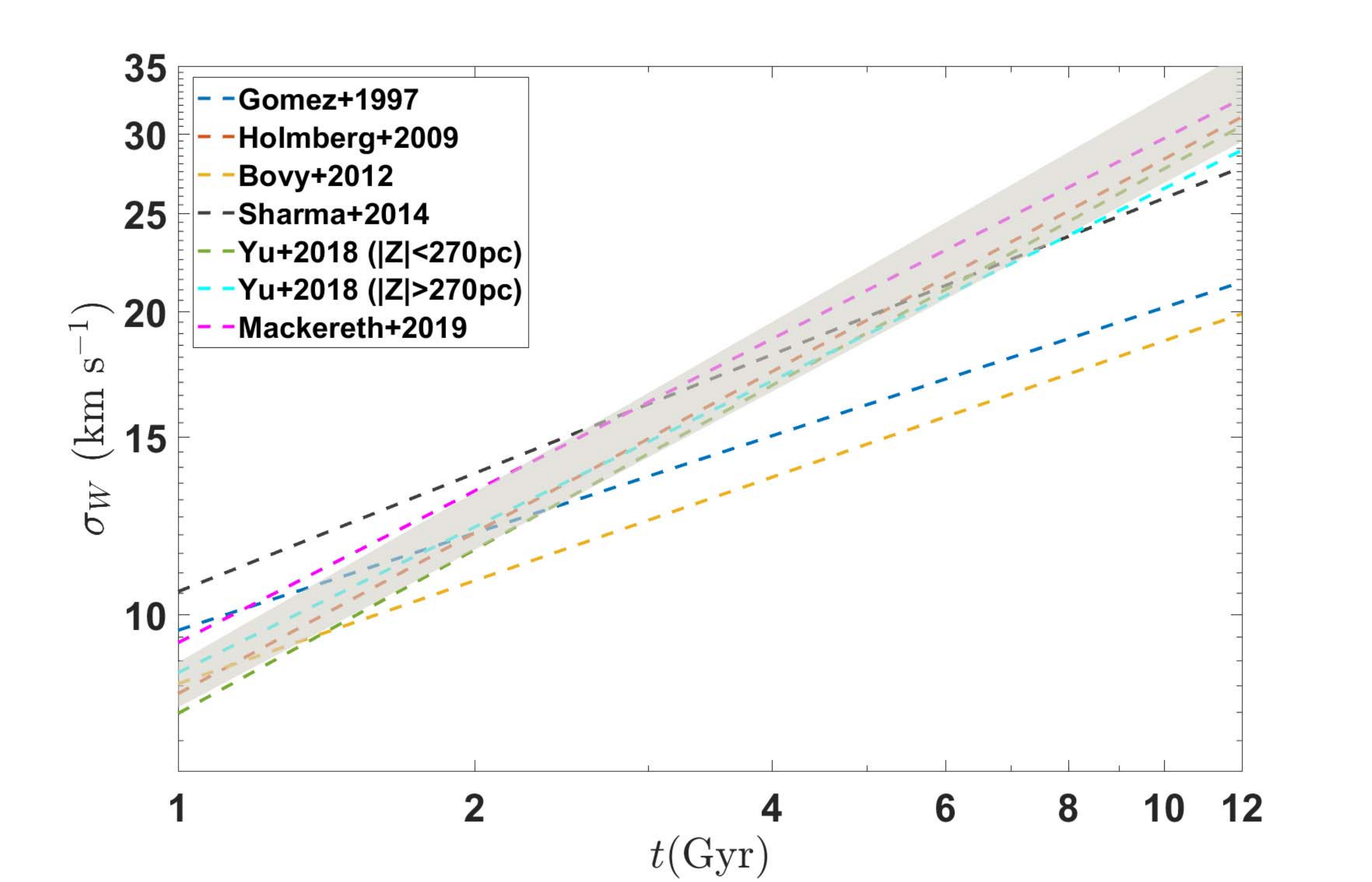}
\caption{{\chen The vertical AVRs from different studies.}
The dashed lines present results from previous studies. The grey region denotes the $1-\sigma$ range of the best fit obtained from our calibration sample. 
\label{figAVRhistory}}
\end{figure}

{\xie Here, we compare our fitted AVR (section \ref{sec.meth.AVR.fit}) to those AVRs fitted with different samples in previous studies.
The result is plotted in Figure \ref{figAVRhistory}.
As can be seen, our result is in good agreement with those of \citet{2009A&A...501..941H, 2018MNRAS.475.1093Y, 2019MNRAS.489..176M}}.
Nevertheless, we note that our result differs significantly from those of  \citet{1997ESASP.402..473B, 2012ApJ...759..131B}, and \citet{2014ApJ...793...51S} on the younger and older ends, respectively.

{\xie All these studies fit their AVRs with different stellar samples.
\citet{1997ESASP.402..621G} used 2,812 stars from the Hipparcos data.
\citet{2009A&A...501..941H} used 2,640 FG main-sequence stars from the Geneva-Copenhagen Survey (GCS) data.
\citet{2012ApJ...759..131B} used 3,365 stars from the APOGEE data.
\citet{2014ApJ...793...51S} used 5,201 stars from the GCS and RAVE data.
\citet{2018MNRAS.475.1093Y} used 3,564 sub-giant/red giant stars from the LAMOST-Gaia data.
\citet{2019MNRAS.489..176M} used 65,719 stars from the APOGEE and Gaia data.}

{\xie Our study used 130,403 stars which were selected from the LAMOST DR4 value added catalog with well constrained kinematics and ages (section \ref{sec.meth.rev.cal}). 
Due to the upgrade of sample size as well as  the quality of star characterizations, the uncertainties of the AVR parameters have been largely reduced in this work. 
To better demonstrate this point, we performed a more detail comparison to the AVR of \cite{2009A&A...501..941H}, which are widely used and very close to our best fit of the AVR (Figure \ref{figAVRhistory}). 
We adopted the same procedures (section \ref{sec.meth.AVR.fit}) to obtain the uncertainties of AVR but using the data of \cite{2009A&A...501..941H}.
}
The results are summarized at the bottom part of Table \ref{tab:AVRparaHT}. 
As can be seen, the typical errors in $k$ and $\beta$ decrease from $\sim 1-3\, \rm km \ s^{-1}$ and $0.05-0.08$ to $\sim 0.5\, \rm km \ s^{-1}$ and $0.02${\xie, respectively}. 
The reduction of uncertainties in AVR parameters leads to significant improvement in the precision of kinematic age (section \ref{sec.kine.age.err}). 

\subsection{Kinematic Age and Uncertainty}
\label{sec.kine.age.err}

{\xie For a group of stars, the typical kinematic age can be derived by using the AVR (solving the Equation \ref{AVR}), which gives}
\begin{equation}
t = \left(\frac{\sigma}{k\rm\, km \ s^{-1}}\right)^{\frac{1}{\beta}}\, \rm Gyr,
\label{kineage}
\end{equation}
By means of error propagation, the {\xie relative} uncertainty of kinematic age can be estimated as :
\begin{equation}
\begin{aligned}
\frac{\Delta t}{t}
        &= \sqrt{{(\frac{\partial \ln{t}}{\partial \beta} \Delta \beta)}^2
        +{(\frac{\partial \ln{t}}{\partial k} \Delta k)}^2+
        {(\frac{\partial \ln{t}}{\partial \sigma} \Delta \sigma)}^2}\\
       & = \sqrt{{\left(\ln \frac{t}{\rm Gyr}\right)^2 \left(\frac{\Delta \beta}{\beta}\right)^2
        +\frac{1}{\beta^2} \left(\frac{\Delta k}{k}\right)^2
        +\frac{1}{\beta^2} \left(\frac{\Delta\sigma}{\sigma}\right)^2}},
\label{kineageerr}
\end{aligned}
\end{equation}
where $\Delta$ represents the absolute uncertainty. {\respondtoZheng For the sake of simplicity, here we assume $k$, $\beta$, and $\sigma$ are independent of each other. In other word, we neglect the  covariances between them.}
It can be seen from the formula: the first term (due to the uncertainty of $\beta$) has positive correlation with the age ($t$), while the latter two terms are independent with age. 
Therefore, the relative uncertainty in age generally increases with age itself.
{\xie For this reason, we set t=1 Gyr and t=10 Gyr to estimate the range of the relative uncertainty of age. }

{\xie Based on Equation \ref{kineageerr}, we analyzed the budget of kinematic age uncertainty derived from AVR.
The results are listed in the left part of Table \ref{tab:AVRparaHT}.
The uncertainties in AVR fitting parameters, i.e., $\Delta k$ and $\Delta \beta$, were calculated as the half width of the 1-sigma interval of $k$ and $\beta$ as listed in Table \ref{tab:AVRpara}.} 
The relative uncertainties in velocity dispersions ($\Delta \sigma/\sigma$) were adopted as the median relative uncertainties of velocity dispersions in our planet host stellar sample, which are 2.0\%, 3.5\%, 4.1\%, and 2.4\% for $\sigma_U, \ \sigma_V, \ \sigma_W$ and $\sigma_{\rm tot}$, respectively.
{\xie Putting all the above uncertainties into Equation \ref{kineageerr}, we obtained the relative age uncertainties $\Delta t/t\sim9.7-14.1\%$, $\Delta t/t\sim11.8-20.1\%$, $\Delta t/t\sim12.3-16.9\%$, and $\Delta t/t\sim9.2-16.0\%$ for $\sigma_U, \ \sigma_V, \ \sigma_W$, and $\sigma_{\rm tot}$, respectively.  
The lower and upper values are calculated by assuming $t = 1$ Gyr and $t= 10$ Gyr respectively in Equation \ref{kineageerr}.}

For comparison, we repeated the above budget calculation but using AVR of \citet[][see the bottom part of Table \ref{tab:AVRpara}]{2009A&A...501..941H}.
The results are listed in the right part of Table \ref{tab:AVRparaHT}.
As can be seen, for AVR from \cite{2009A&A...501..941H}, the uncertainties of $k$, $\beta$ are much larger and thus dominant, leading to a much larger (by a factor of $\sim 3$) uncertainty in the derived kinematic age.

As discussed in section \ref{sec.meth.AVR.diffcom}, AVRs are not significant for halo and Hercules stream, therefore  we suggest that this method to obtain kinematic age is {\xie only} suitable for stars {\xie belonging to the {\respondtoxiang Galactic} disk components.}

\begin{table}[!t]
\centering
\caption{The typical relative uncertainties of parameters and kinematic ages derived by the Age-Velocity Relations using data from \cite{2009A&A...501..941H} and this work. }
{\footnotesize
\label{tab:AVRparaHT}
\begin{tabular}{c|c|cc} \hline
&  & This work  & Holmberg et al. 2009 \\ \hline
\multirow{4}{*}{$U$} & $\Delta k/k$   & 2.6\% & 10.1\% \\
 & $\Delta \beta/\beta$ & 4.4\% & 20.5\% \\
 & $\Delta \sigma/\sigma$ & 2.0\% &  2.0\%\\ \cline{2-4}
 & $\Delta t/t$ & $9.7\%-14.1\%$ & $31.6\%-64.1\%$  \\ \hline
\multirow{4}{*}{$V$} & $\Delta k/k$   & 3.7\% & 9.4\% \\
 & $\Delta \beta/\beta$ & 4.1\% & 13.8\% \\
 & $\Delta \sigma/\sigma$ & 3.5\% &  3.5\%\\ \cline{2-4}
 & $\Delta t/t$ & $11.8\%-20.1\%$ & $23.32\%-44.4\%$  \\ \hline
\multirow{4}{*}{$W$} & $\Delta k/k$   & 5.2\% & 14.1\% \\
 & $\Delta \beta/\beta$ & 3.7\% & 12.3\% \\
 & $\Delta \sigma/\sigma$ & 4.1\% &  4.1\%\\ \cline{2-4}
 & $\Delta t/t$ & $12.3\%-16.9\%$ & $27.2\%-45.3\%$  \\ \hline 
 \multirow{4}{*}{$V_{\rm tot}$} & $\Delta k/k$   & 2.8\% & 10.7\% \\
 & $\Delta \beta/\beta$ & 5.0\% & 15.0\% \\
 & $\Delta \sigma/\sigma$ & 2.4\% & 2.4\%\\ \cline{2-4}
 & $\Delta t/t$ & $9.2\%-16\%$ & $27.4\%-54.1\%$  \\ \hline

\end{tabular}}
\end{table}


\section{Application to Planet Host Stars}
\label{sec.res}
{\chen In this section, we apply the above revised kinematic method (section \ref{sec.meth.classify}) and AVR (section \ref{sec.meth.avr}) to a sample of planet host stars (section \ref{sec.samp}), providing a catalog of their kinematic properties (section \ref{sec.res.cat}), with focus on the Galactic components (section \ref{sec.res.com}) and kinematic ages (section \ref{sec.res.age}).}

\subsection{Data samples}
\label{sec.samp}
{\xie
This subsection describes how we constructed the planet host sample for further kinematic characterizing. 
}

\subsubsection{initializing planet host sample from EA}
{\xie We initialized our planet host sample with the confirmed planets table and the Kepler DR 25 catalog from EA. }
The Kepler catalog contains 8,054 Kepler Objects of Interest (KOIs) in DR25. 
Here, we excluded KOIs flagged by `False Positive' \citep[FAP,][]{2018ApJS..235...38T}, leaving 4,034 planets (candidates) around 3,069 stars.
{\xie Besides Kepler, there are 1,728 non-Kepler planets flagged by `Confirmed' around 1,387 stars.}
{\xie We also removed potential binaries because additional motions caused by binary orbits could affect the results of kinematic characterization.
Specifically, for Kepler planet host stars, we eliminated stars with Gaia DR2 re-normalized unit-weight error
(RUWE) $>1.2$ \citep{2018AJ....156..195R,2020AJ....160..108B}.
For non-Kepler planet host stars, we excluded those with $\rm pl\_cbflag \ne 0$ (flag indicating whether the planet orbiting a binary flag, 0 for no) in EA. 
In total, we are left with 4,126 stars hosting 5,331 planet candidates in our initial sample.}




\subsubsection{obtaining five astrometric parameters from Gaia}
\label{sec.samp.5ast}
{\xie Next, we cross-matched our initial planet host sample with Gaia to obtain astrometric parameters.}
The second Gaia data release (DR2, \cite{2018A&A...616A...1G}) includes five astrometric parameters: positions on the sky $(\alpha, \delta)$, parallaxes, and proper motions ($\mu_{\alpha}, \mu_{\delta}$) for more than 1.3 billion stars, with a limiting magnitude of G $ = 21$ and a bright limit of G $\approx 3$. {\xie The cross-matching was done} by using the X match service of the Centre de Donnees astronomiques de Strasbourg (CDS, http://cdsxmatch.u-strasbg.fr). 
{\xie The separation limit of the cross-matching was chosen as where the distribution of separations displayed a minimum,  $\sim1.5$ arcseconds.} 
Besides the separation condition,  we also made a magnitude cut to ensure that the matched stars are of similar brightness. 
{\xie The magnitude limit was set by inspecting the distribution of magnitude difference, which is 2 mag in Gaia G mag.} 
{\xie If multiple matches satisfied these two criteria, we kept the one with the smallest angular separation.}
Finally, we obtained 5,069 planets around 3,912 stars.


\begin{table*}[!ht]
\centering
\caption{Compositions of our planet host sample.}
{\footnotesize
\label{tab:planetnumberprocedure}
\linespread{1.8}
\begin{tabular}{p{4.5cm} p{1.6cm}<{\centering} p{1.6cm}<{\centering} p{1.6cm}<{\centering} p{1.6cm}<{\centering} p{1.6cm}<{\centering} p{1.6cm}<{\centering}} \hline
      & \multicolumn{3}{c}{--------------~~Space-based~~--------------} & \multicolumn{3}{c}{--------------~~Ground-based~~--------------}   \\ 
     & Kepler & K2 & CoRot & RV & Transit & Others\\ \hline \hline 
    \multicolumn{7}{c}{section 4.1.1: Initializing Planet Host Sample from EA} \\ \hline
\multirow{2}{*}{Without FAP\& binary: 4126 (5331)} & 2737& 283& 29& 590& 366 &121 \\ 
         & (3620) & (389)  & (31) & (775) & (384) & (132) \\ \hline \hline
\multicolumn{7}{c}{section 4.1.2: Obtaining Five Astrometric Parameters from Gaia} \\ \hline
\linespread{0.8}
\multirow{2}{*}{With Gaia Astronomy: 3912 (5069)}  & 2571 & 282&  29& 568& 364 &98 \\
  & (3409) & (388) & (31) & (752) & (382) & (107) \\ \hline \hline
\multicolumn{7}{c}{section 4.1.3: Obtaining RV from Various Sources} \\ \hline
\multirow{2}{*}{APOGEE: 692 (956)}  & 628 & 30 & 2 & 21 &9 & 2 \\
  &  (874) & (42) & (3) & (24) & (11) & (2) \\ 
\multirow{2}{*}{RAVE: 30 (37)}  & 0 & 4 & 0 & 6 & 20 & 0 \\
  & (0) & (6) & (0) & (10) & (21) & (0) \\
\multirow{2}{*}{Gaia: 1143 (1479)}  & 268 & 145 & 3 & 497 & 223  & 7 \\
 & (371) & (204) & (4) & (659) & (234)  & (7) \\
\multirow{2}{*}{LAMOST: 1059 (1421)} & 951 & 38 & 0  & 24 & 45  & 1 \\
 & (1292) & (54) & (0) & (26) & (47)  & (2) \\
\multirow{2}{*}{EA: 1303 (1737)}  & 215 & 185 & 20 & 529 & 338  & 16 \\
&  (371)  & (266) & (23) & (709) & (351)  & (17) \\ \hline \hline
\multicolumn{7}{c}{section 4.1.4: Finalizing the Planet Host Sample} \\ \hline
{\chen \multirow{2}{*}{Combined: 2174 (2872)}} & {\chen 1134} & {\chen 179} & {\chen 20} & {\chen 516} & {\chen 306} & {\chen 19} \\
&  {\chen (1562)} & {\chen (249)} & {\chen (22)} & {\chen (699)} & {\chen (319)} & {\chen (21)} \\\hline \hline
\end{tabular}}
\flushleft
{\scriptsize
  The numbers without and with brackets represent that of the stars and planets during the process of sample selection in section \ref{sec.samp}.}
\end{table*}

\subsubsection{obtaining RV from various sources}
\label{sec.samp.rv}
We obtained radial velocities from the following five catalogs: the APOGEE data release (DR) 16 catalog, the LAMOST DR4 value-added catalog, the RAVE DR5 catalog, Gaia, and EA. 

\begin{itemize}
\item {\textbf{APOGEE}}
{\xie The APOGEE DR16  provides a catalog of 437,485 unique stars, which contains the information of radial velocity (RV), effective temperature ($T_{\rm eff}$), surface gravity ($\log g$), and chemical abundance (e.g,. $\rm [Fe/H]$ and $\rm [\alpha/Fe]$)  \citep{2020ApJS..249....3A}.
}
We cross-matched it with {\xie the planet host sample obtained in section \ref{sec.samp.5ast}.}
Here, we applied the following quality control cuts: (1) $\rm STARFLAG = 0$ to only select star with no warnings on the observation; (2) $\rm ASPCAPFLAG = 0$ to ensure parameters have converged and no warning; {\xie and} (3) $\rm SNR>80$ to ensure high SNR \citep{2018AJ....156..125H}, {\xie leaving 692 stars hosting 956 planets.}

\item {\textbf{RAVE}}
The fifth data release (DR5) {\xie of RAVE} provides {\xie radial velocities} with a precision of $\sim 1.5 \rm km\ s^{-1}$ and physics properties ($T_{\rm eff}$, $\log g$, $\rm [Fe/H]$, etc.) from a magnitude-limited $(9<I<12)$ survey for 457,588 randomly-selected stars in the southern hemisphere \citep{2017AJ....153...75K}.
{\xie To cross-match it with the planet host sample obtained in section \ref{sec.samp.5ast},  we applied the following quality cuts:} (1) $\rm Algo\_Conv_K=0$ to ensure that the
stellar parameter pipeline has converged; (2) $\rm SNR>40$; (3) spectroscopic morphological flags (c1, c2, c3) are n; (4) $\rm Alpha\_C>-9.99$  \citep{2017AJ....153...75K}. 
{\xie The RAVE DR5 contains stars only in a range of declination from $-88$ to $+28$ deg,
thus the Kepler field which ranges from  36 to 53 deg in declination is not covered. 
Therefore, the above cross-matching with RAVE returned only 30 stars hosting 37 non-Kepler planets.}

\item {\textbf{Gaia}} 
Gaia DR2 also includes radial velocities for more than 7.2 million stars with a {\xie magnitude range of G$\sim 4-13$} and a $T_{\rm eff}$ range of about 3550 to 6900 K. 
The quality control cut is set as that {\xie the ratio of radial velocity and its uncertainty should be  larger than 3.
After cross-matching with the planet host sample obtained in section \ref{sec.samp.5ast}, we obtained 1,143 stars hosting 1,479 planets (268 of them are stars with 371 planets in the Kepler field).}

\item {\textbf{LAMOST}}
The LAMOST survey has several components focusing on different {\respondtoxiang Galactic} aspects, e.g., the {\respondtoxiang Galactic} halo  \citep{2012RAA....12..735D}, stellar clusters  \citep{2013IAUS..292..105H}, the {\respondtoxiang Galactic} anticenter (LSS-GAC; \citet{2014IAUS..298..310L}), the Kepler fields  \citep{2015ApJS..220...19D}, and etc.
The LAMOST DR4 value-added catalog  $\footnote{http://dr4.lamost.org/v2/doc/vac}$ contains parameters derived from a total of 6.5 million stellar spectra for 4.4 million unique stars  \citep{2017MNRAS.467.1890X}. 
RVs, $T_{\rm eff}$, $\log g$, and $\rm [Fe/H]$ have been deduced using both the official LAMOST Stellar parameter Pipeline (LASP; \cite{2011RAA....11..924W}) and the LAMOST Stellar Parameter Pipeline at Peking University (LSP3; \cite{2015MNRAS.448..822X}). 
{\xie The typical uncertainties for RVs, $T_{\rm eff}$, $\log g$, and $\rm [Fe/H]$} are 5.0 $\rm km \ s^{-1}$, 150 K, 0.25 dex, and 0.15 dex respectively. 
{\xie After applying a quality cut of $\rm SNR>10$ and cross-matching with planet host sample obtained in section \ref{sec.samp.5ast}, we obtained 1,059 stars hosting 1,421 planets. The majority (951) are stars with Kepler planets (1,292).}

\item {\textbf{EA}}
{\xie The NASA Exoplanet Archive (EA) also reports RVs for a portion of stars. 
We thus cross-matched these stars with planet host sample obtained in section \ref{sec.samp.5ast}.
The quality control cut is also set as that the ratio of radial velocity and its uncertainty should be larger than 3, which yields 1,303 stars hosting 1,737 planets. 
The majority (1,088) are stars with non-Kepler planets (1,366) from various ground based RV and transit surveys.
Note, most RV data in EA are collected from various literatures and thus are inhomogeneous.} 
\end{itemize}

\subsubsection{finalizing the planet host sample}
%



\begin{figure}[!t]
\centering
\includegraphics[width=0.5\textwidth]{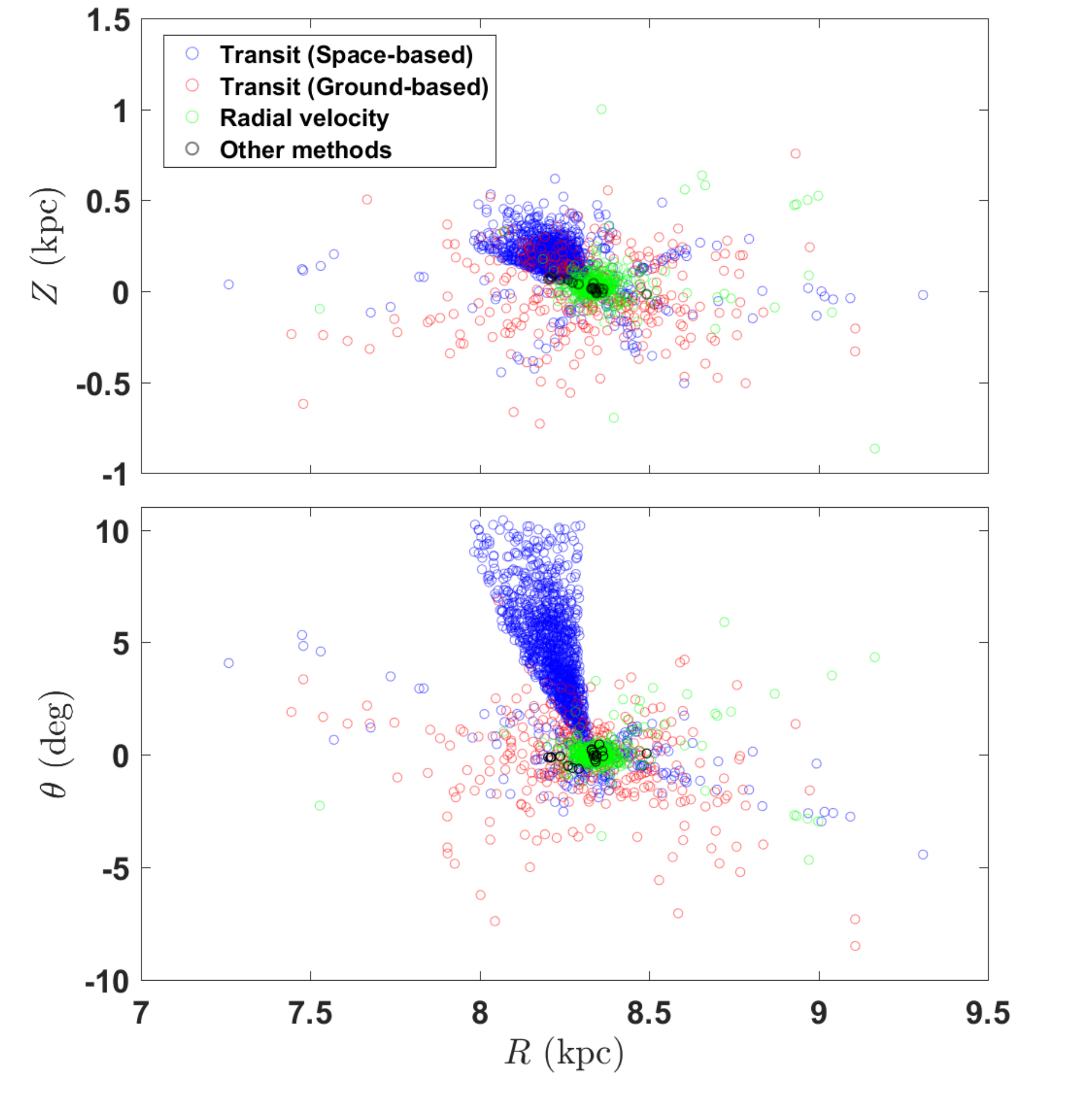}
\caption{\chen Galactocentric radius ($R$) vs. height ($Z$, top panel)  and angle ($\theta$, bottom panel)for the combined planet host sample.
The diagram is colour-coded to represent different discovery methods and facilities.
(8.34 kpc, 0, 0.027 kpc) marks the position of Sun. 
\label{figRtheZstars}}
\end{figure}

{\xie We finalize the planet host sample by combining various matched samples in section \ref{sec.samp.rv}.
For stars with multiple RV measurements from different sources, we take the order of precedence as APOGEE, RAVE, Gaia, LAMOST, then EA.  
This generally follows the order of spectral resolution and thus the RV uncertainty. 
Here we set the EA as the lowest priority because the most RVs from EA are collected from various sources, which are inhomogeneous.
{\chen For the sake of reliability, we exclude the stars if the differences in their RVs from different sources are larger than three times of the uncertainties.
We also apply the same cut as in the calibration sample, i.e. distance $< 1.54$ kpc, corresponding to ($7< R < 10$ kpc, $\theta<10$ deg, and $|Z|<1.5$ kpc).
Finally, we obtain a sample of 2,174 stars hosting 2,872 planets.}
In Table \ref{tab:planetnumberprocedure}, we summarize the composition of the sample after each step mentioned above.
Figure \ref{figRtheZstars} {\xie shows} the location of stars in our planet host {\xie sample}. 
}

\subsection{ A Catalog of Planet Hosts with Kinematic Characterizations}
\label{sec.res.cat}
{\xie Applying the methods described in section \ref{sec.meth.classify} and section \ref{sec.meth.avr} to the planet host sample (section \ref{sec.samp}), we obtained a catalog (Table \ref{tab:starcatalog}) of {\chen 2,174} planet hosts with kinematic characterizations, e.g., Galactocentric velocities to the LSR ($U_{\rm LSR}$,$V_{\rm LSR}$, and $W_{\rm LSR}$) and the relative membership probabilities between different {\respondtoxiang Galactic} components ($TD/D$, $TD/H$, $Her/D$, and $Her/TD$)}.
For the sake of completeness, we also put in the catalog the stellar parameters that used during the process of our kinematic characterization (e.g., parallax, proper motion, and RV) and other basic stellar parameters (e.g., $T_{\rm eff}$, $\log g$, $\rm [Fe/H]$, and $\rm [\alpha/Fe]$).
As mentioned before,  for stars with multiple sources of RV, the order of precedence follows the order of spectral resolution and thus the uncertainty, i.e. as APOGEE, RAVE, Gaia, LAMOST then EA. 
While for other stellar parameters, because the estimates of stellar parameters from APOGEE are only reliable for relatively cool stars $4000<T_{\rm eff}<5500$ K \citep{2018AJ....156..125H},  we took APOGEE as the lowest order of precedence here instead.
In what follows, we conduct some analyses on this catalog.


\begin{table*}[!ht]
\centering
\caption{\chen The numbers of stars (planets) of our planet host sample in different Galactic components}
{\footnotesize
\label{tab:starnumber}
\begin{tabular}{ll|cccccc} \hline
     &     & Total & Thin disk & Thick disk  & Hercules & Halo & In between \\ \hline
    \multicolumn{2}{c|}{Radial Velocity} & 516 (699) & 440 (602) & 25 (33)  & 20 (30) & 1 (1) & 30 (33) \\ \hline
    \multirow{4}{*}{Transit} & Kepler & 1134 (1562) & 982 (1363) & 68 (89)  & 15 (20) & 0 (0) & 69 (90) \\
                             & K2 & 179 (249) & 156 (221) & 8 (9)  & 5 (9) & 0 (0) & 10 (10) \\
                             & CoRoT & 20 (22) & 19 (21) & 1 (1) & 0 (0) & 0 (0) & 0 (0) \\
                             & Ground-based & 306 (319) & 278 (290) & 12 (12)  & 5 (5) & 0 & 11 (12) \\ \hline
    \multicolumn{2}{c|}{Other methods} & 19 (21) & 19 (21) & 0 (0) & 0 (0) & 0 (0) & 0 (0)\\ \hline
    \multicolumn{2}{c|}{All} & 2174 (2872) & 1894 (2518) & 114 (144) & 45 (64) & 1 (1) & 120 (145)\\ \hline
\end{tabular}}
\flushleft
{\scriptsize
 The numbers without and with the brackets represent that of stars and planets.}
\end{table*}        
   
\begin{table*}[!ht]
\centering
\caption{\chen {The summary of Galactic velocities and chemical abundances for the combined sample of planet host stars}}
{\footnotesize
\label{tab:starVCA}
\begin{tabular}{l|cccccc} \hline
                & \multicolumn{2}{c}{---------~~$V_{\rm tot} \ \rm (kms^{-1})$~~---------} & \multicolumn{2}{c}{---------~~$\rm [Fe/H]\  \rm (dex)$~~---------} & \multicolumn{2}{c}{---------~~$\rm [\alpha/Fe] \ \rm dex$~~---------}    \\ 
               &  value & $1-\sigma$ interval &  value & $1-\sigma$ interval &  value & $1-\sigma$ interval \\ \hline
 Thin disk  & 34.8 & (19.1, 56.0) & $0.00$ & $(-0.16,0.21)$ &  $0.01$ & $(-0.19,0.19)$   \\
    Thick disk  & 97.8 & (80.9, 124.4) & $-0.20$ & $(-0.47,0.11)$ &  $0.15$ & $(0.03,0.26)$  \\
    Halo  & 282.6 & NA & $-0.89$ & NA & NA & NA \\
    Hercules  & 73.5 & (62.7, 91.1) & $-0.05$ & $(-0.40, 0.13)$ &  $0.07$ & $(-0.08,0.17)$ \\ \hline
 
\end{tabular}}
\end{table*}

\subsection{{\respondtoxiang Galactic} Components of Planet Hosts}
\label{sec.res.com}
{\xie With the derived relative membership probabilities between different {\respondtoxiang Galactic} components ($TD/D$, $TD/H$, $Her/D$, and $Her/TD$ in Table \ref{tab:starcatalog}), we then classify the {\chen 2,174} planet host stars into four {\respondtoxiang Galactic} components, i.e., thin disk, thick disk, Hercules stream, and halo following the method as mentioned in section \ref{sec.meth.class}.  
For stars not belonging to the above four components, following \citet{2014A&A...562A..71B}, we classify them into a category dubbed as `in between'.

The results of the classification are summarized in
Table \ref{tab:starnumber}, which lists the numbers of stars in different categories.}
{\xie As can be seen, about {\chen 87.1\% (1,894/2,174)} of stars in our sample are in thin disk and about {\chen 5.2\% (114/2,174)} stars are in thick disk. }
{\chen 45} stars in the planet host sample are affiliated to Hercules stream, which has been speculated to have a dynamical origin in the inner parts of the Galaxy and then  kinematically heated by the central bar  \citep{2005A&A...430..165F,2007ApJ...655L..89B}. 
The fraction of halo stars is {\chen $\sim$ 0.05\% (1/2,174)} and there are another {\chen $\sim$ 5.5\% (120/2,174)} belonging to the `in between' category. 
In Table \ref{tab:starnumber}, we also divide planet host stars according to the method that discovered the planets. 
{ In general, we find that, first, for transiting planet hosts, those observed from space-base facilities have a higher fraction of thick disk {\chen (5.8\%, 77/1,333)} than those observed from ground-base {\chen (3.9\%, 12/306)}.
Second, for ground-based planet hosts, RV planet hosts have a higher fraction of thick disk fraction {\chen (4.8\%, 25/516)} than those transiting planet hosts {\chen (3.9\%, 12/306)}.}

\begin{figure}[!t]
\centering
\includegraphics[width=0.5\textwidth]{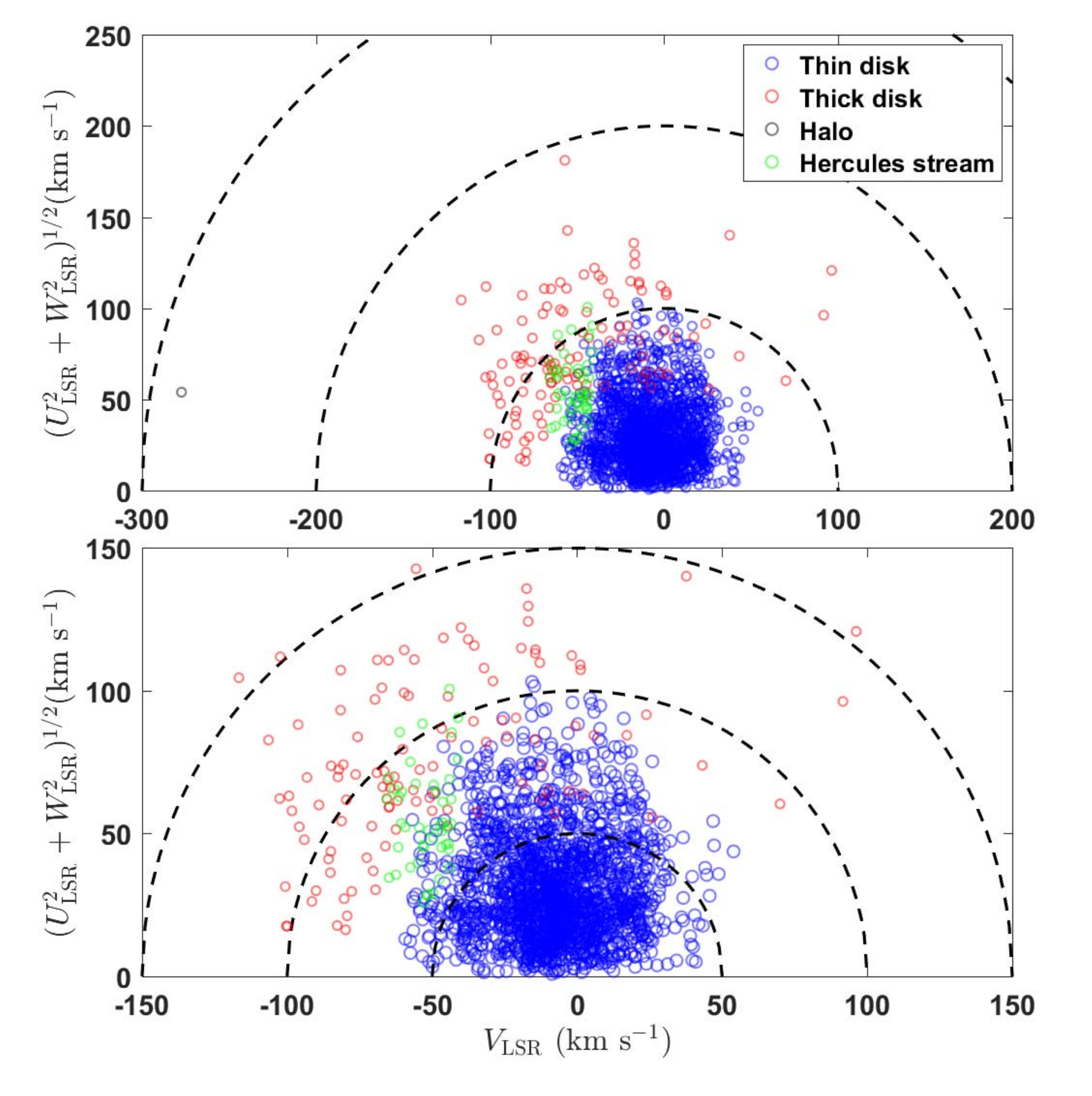}
\caption{\chen The Toomre diagram of combined planet host sample for different Galactic components. The top panel shows the full range of velocities while the bottom panel zooms in the region where the majority of the data points are located.
The diagrams are colour-coded to represent different components. 
Dashed lines show constant values of the total Galactic velocity $V_{\rm tot} = (U_{\rm LSR}^2+V_{\rm LSR}^2+W_{\rm LSR}^2)^{1/2}$, in steps of 100 and 50 $\rm km \ s^{-1}$ respectively in the two panels.
\label{figGLKUVWpopu}}
\end{figure}

\begin{figure*}[!t]
\centering
\includegraphics[width=\textwidth]{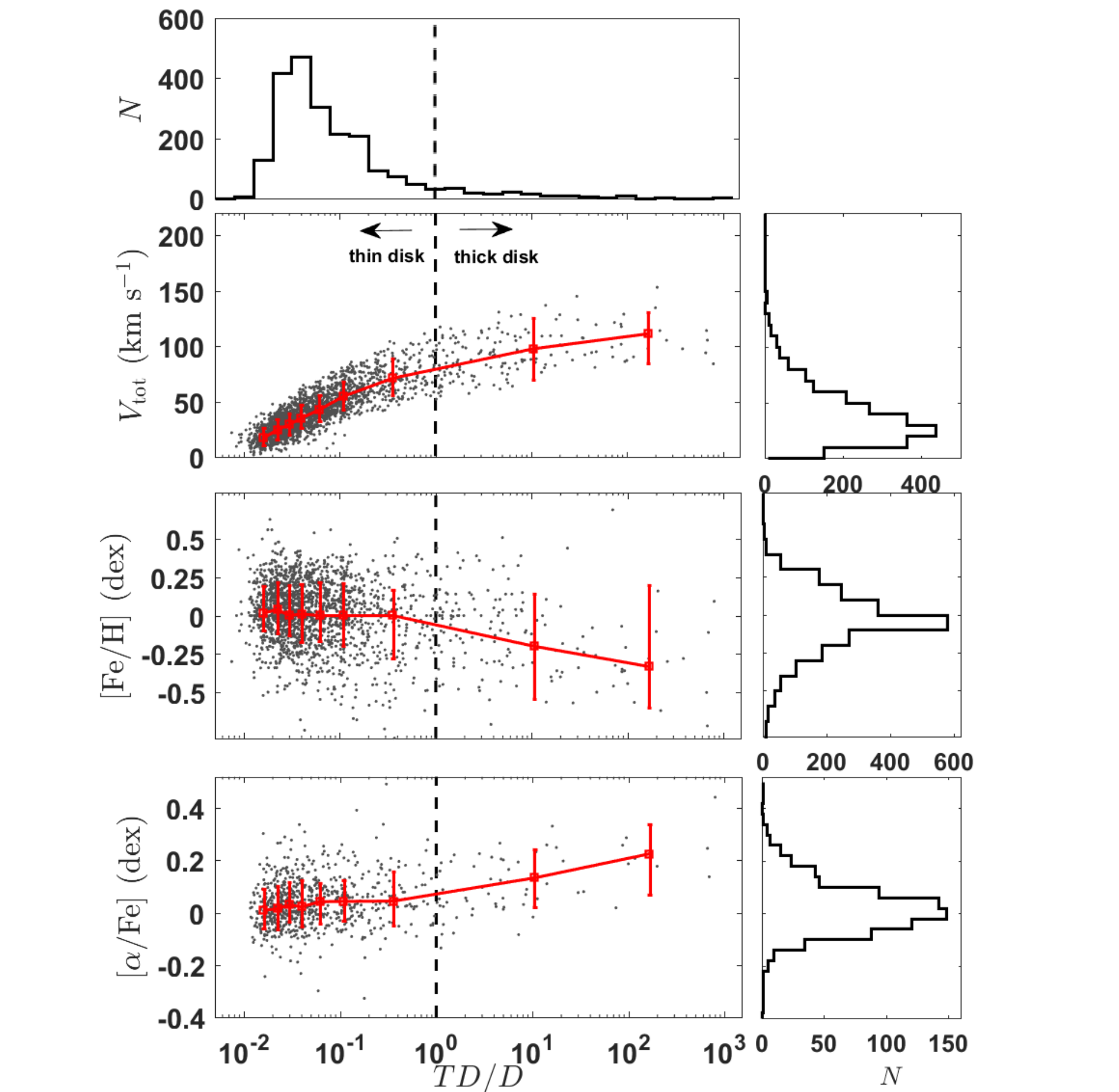}
\caption{\chen Top panel: The relative probabilities for the thick-disk-to-thin-disk $TD/D$ vs. total velocity $V_{\rm tot}$; Middle pannel: $TD/D$ vs. $\rm [Fe/H]$; Bottom panel: $TD/D$ vs. $\rm [\alpha/Fe]$.
Medians and $1-\sigma$ dispersions are marked in the red color.
Histograms of $\rm V_{\rm tot}$, $\rm [Fe/H]$ and $\rm [\alpha/Fe]$ are shown in the right panels. Histogram of $TD/D$ is displayed in the toppest.
The vertical dashed lines represent where $TD/D = 1$.
\label{figTDDVtotFealpha}}
\end{figure*}

We plot the Toomre diagram of the planet host stars in Figure \ref{figGLKUVWpopu}. 
As can be seen, the boundaries of different components are well consistent with the results of previous works \citep[e.g][]{2014A&A...562A..71B}.
In specific, most stars with low velocities ($V_{\rm tot}\lesssim 50 \  \rm km\ s^{-1}$) are in the thin disk, while those with moderate velocities ($V_{\rm tot}\sim70-180 \ \rm km\ s^{-1}$) are mainly in thick disk. 
The velocity of the only halo star is larger than 220 $\rm km \ s^{-1}$.

We summarized the median values of velocities and chemical abundances for different component in Table \ref{tab:starVCA}. 
As expected, the halo star has the highest {\respondtoxiang Galactic} velocity and poorest $\rm [Fe/H]$, and the thick disk stars are kinematically hotter, metal-poorer ($\sim 0.2$ dex) and $\rm \alpha$-richer ($\sim 0.1$ dex) than the thin disk stars.
The Hercules stream has velocities and chemical abundances which are between those of thin and thick disk stars. 
{\chen We have only one halo star, which does not have $\rm [\alpha/Fe]$ measurement and thus the median and $1-\sigma$ interval of $\rm [\alpha/Fe]$ are not provided.}
In Figure \ref{figTDDVtotFealpha}, we plot the total velocity $V_{\rm tot}$, $\rm [Fe/H]$, and $\rm [\alpha/Fe]$ as a function of $TD/D$. As expected, $V_{\rm tot}$, $\rm [\alpha/Fe]$ increase with $TD/D$, while $\rm [Fe/H]$ decreases with $TD/D$.

{\chen Kapteyn's star (or GJ 191, HD 33793) is the only halo star in our sample. This star has been also identified as the closest halo star to the Sun at a distance of only 3.93 pc with Hipparcos data \citep{2007A&A...474..653V}. 
It is a 11.5 Gyr old M0 type star with a temperature of 3,550 K and estimeated mass of 0.28 solar mass \citep{2010AJ....139..636W,2014MNRAS.443L..89A}.
Interestingly, it is orbited by a confirmed super-Earth (Kapteyn c) with a period of 121.5 days and a candidate super-Earth (Kapteyn b) with a period of 48.6 days.
Furthermore, Kapteyn b lies within the habitable zone \citep{2014MNRAS.443L..89A} and could probably support life at the present stellar activity level \citep{2016ApJ...821...81G}. If confirmed, Kapteyn b will become the oldest habitable planet known so far.
The existence of such a multiple planetary system around a halo star may provide peculiar insights into the planetary formation and evolution at the early time of the Milky way. }

\begin{figure*}[!t]
\centering
\includegraphics[width=\textwidth]{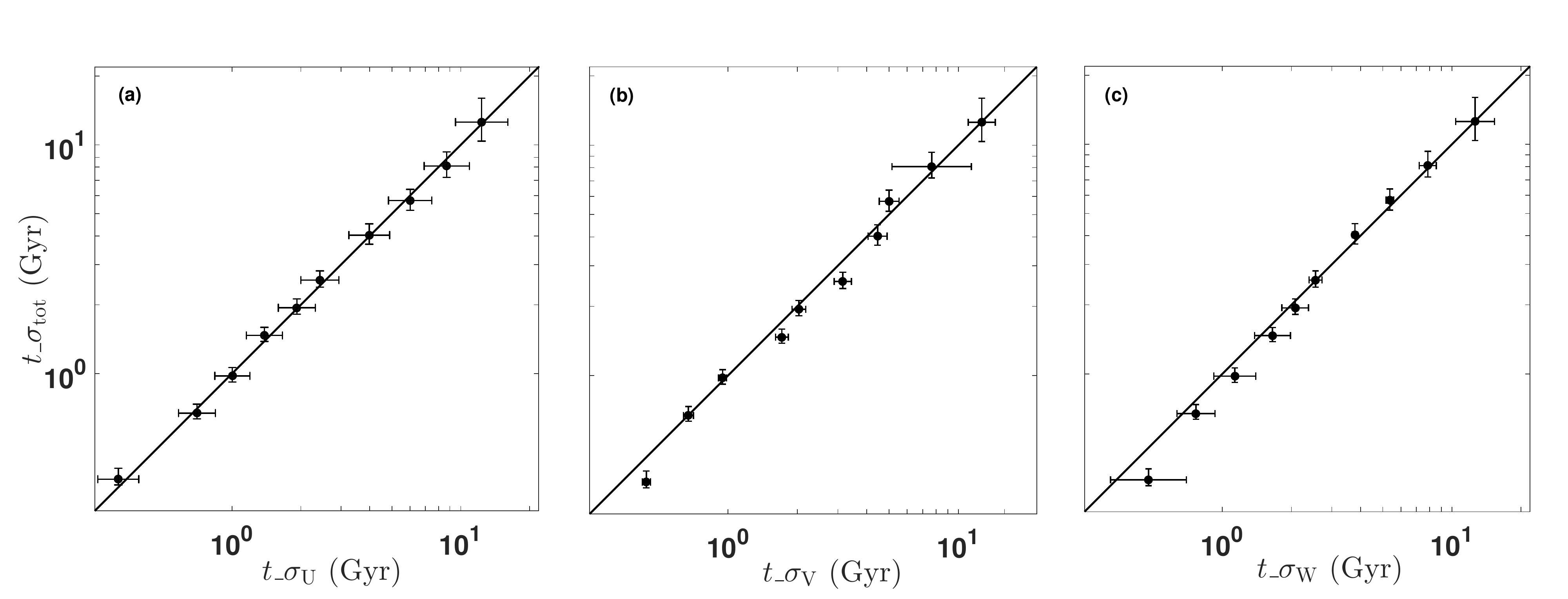}
\caption{\chen Comparisons of the kinematic ages calculated from dispersions of  $V_{\rm tot}$ with ages from $U_{\rm LSR}$ (a), $V_{\rm LSR}$ (b) and $W_{\rm LSR}$ (c) for the planet host sample. The black line in each panel indicates where the horizontal and vertical coordinates are equal to each other. 
\label{figAgeUVWVtot}}
\end{figure*}

\subsection{Kinematic Ages of Planet Hosts}
\label{sec.res.age}
In this section, by using our catalogs of kinematic properties for planet host star, we divided planetary host stars into various groups to study their kinematic ages.
{\xie For each group, we calculated the velocity dispersion to derive the corresponding kinematic age from Equation \ref{kineage}.
To access the uncertainties of the kinematic ages, we took a Monte Carlo method by resampling the AVR parameters ($k$ and $\beta$) and velocity dispersion ($
\sigma$) based on their uncertainties.
For $k$ and $\beta$, their values and uncertainties were adopted from Table \ref{tab:AVRpara}.
For $\sigma$, its value and uncertainty were calculated by resampling each star's {\respondtoxiang Galactic} velocities {\respondtoxiang from a normal distribution given its value and uncertainty.}
Finally, the age uncertainty was set as the 50$\pm$34.1 percentiles in the resampled age distribution.}

\begin{figure}[!t]
\centering
\includegraphics[width=0.5\textwidth]{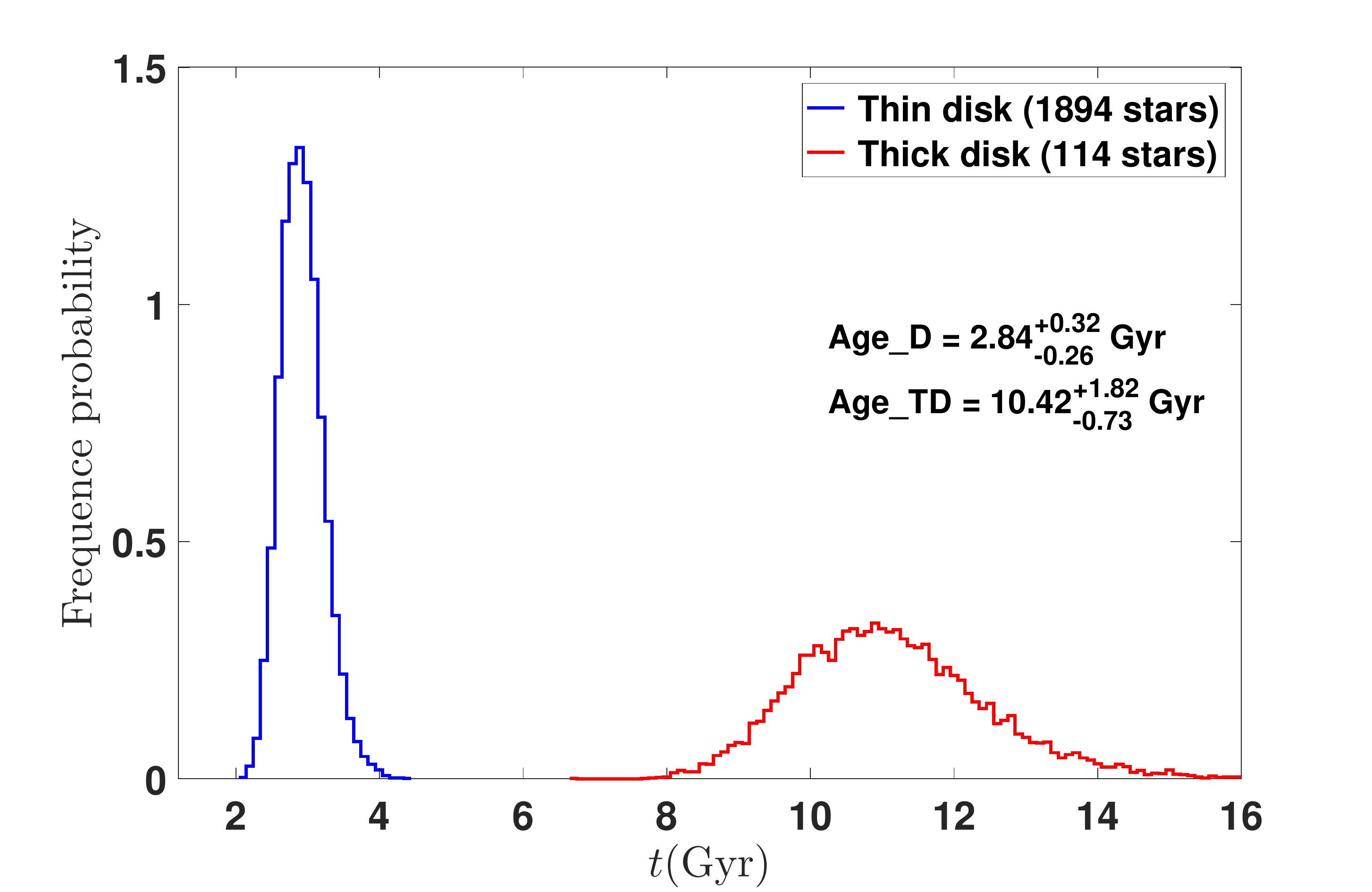}
\caption{\chen Distributions of the kinematic age $t \ (\rm Gyr)$ for 1,894 planet host stars in the thin disk (blue) and 114 planet host stars in the thick disk (red).
\label{figAgedisk}}
\end{figure}

\begin{figure}[!t]
\centering
\includegraphics[width=0.5\textwidth]{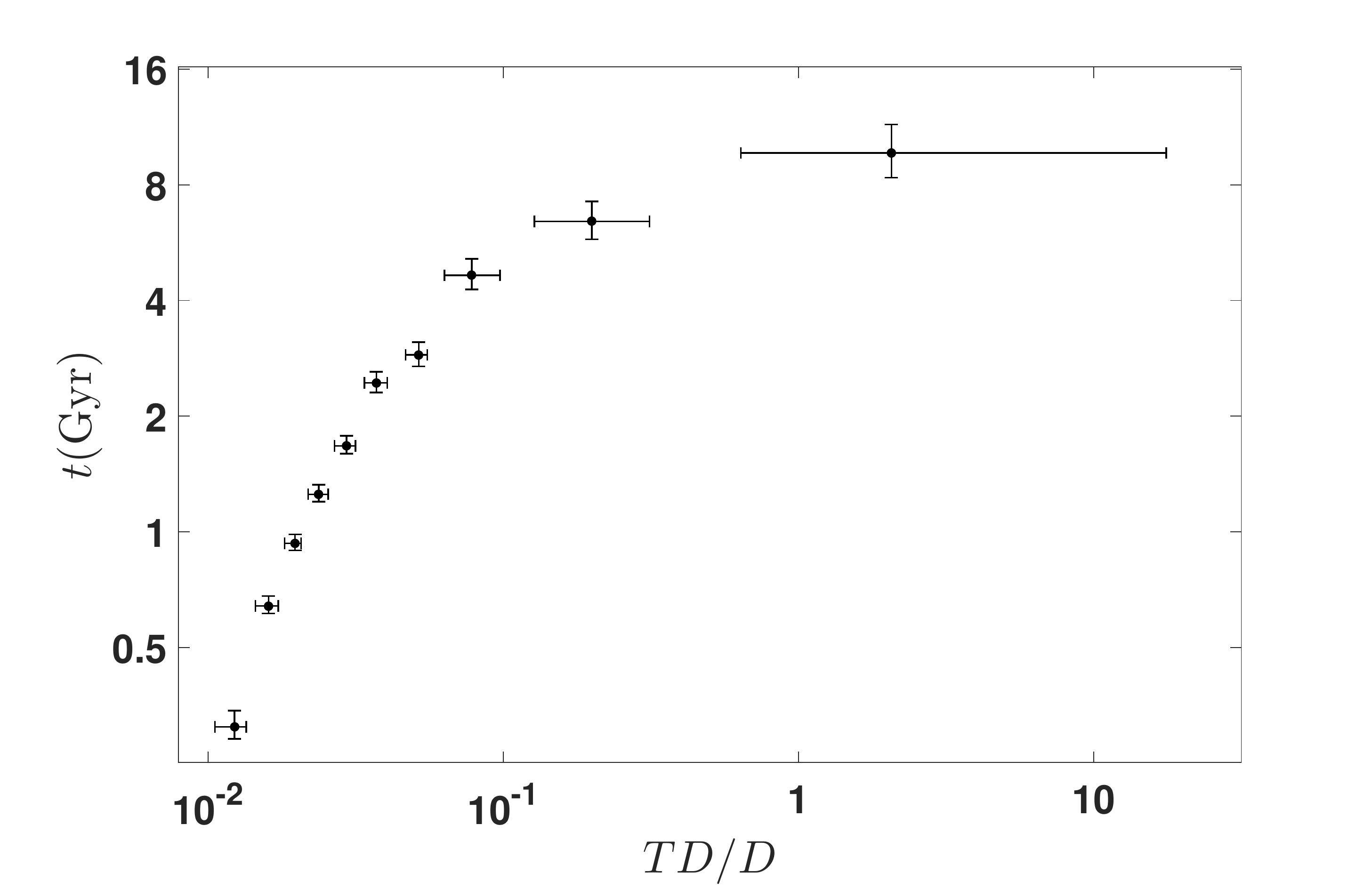}
\caption{\chen The kinematic age $t \ (\rm Gyr)$ vs. the relative probability, $TD/D$ ( thick-disk-over-thin-disk) for our planet host sample.
\label{figAgeTDD}}
\end{figure}

\subsubsection{kinematic ages derived with \\ total velocity and velocity components}
{\xie As shown in Figure \ref{fAVR} of section \ref{sec.meth.avr}, the dispersions of velocity components, i.e., $\sigma_{\rm U}$, $\sigma_{\rm V}$, $\sigma_{\rm W}$, and total velocity $\sigma_{\rm tot}$ all increase with age and fit well with power-law functions. }
{\xie Nevertheless, the fitting parameters ($k$ and $\beta$ in Table \ref{tab:AVRpara}) are different for different velocity components, which in turn could give different kinematic ages.
Here, we compare different kinematic ages calculated from the total velocity and different velocity components.}
{\xie To do this, we first sorted the plant host sample by the relative probabilities for the  thick-disk-to-thin-disk, i.e., $TD/D$. 
Next, according to $TD/D$,} we divided the planet host sample into 10 bins with approximately equal size {\chen ($\sim$ 217)}. 
{\xie 
Then, for each bin, we calculated kinematic ages using $\sigma_{\rm U}$, $\sigma_{\rm V}$, $\sigma_{\rm W}$, and $\sigma_{\rm tot}$.
Finally, we compare these kinematic ages in Figure \ref{figAgeUVWVtot}. 
As can be seen, kinematic ages derived using different velocities are well consistent with each other.
To further see the differences quantitatively, we fit the relations between age derived from $\sigma_{\rm tot}$ and ages from $\sigma_{\rm U}$, $\sigma_{\rm V}$, and $\sigma_{\rm W}$.
{\chen The results are:
\begin{equation}
\begin{aligned}
 &t\_\sigma_{\rm U} =  0.97^{+0.07}_{-0.12} \times t\_\sigma_{\rm tot},  \\
 &t\_\sigma_{\rm V}  =  0.94^{+0.07}_{-0.11} \times t\_\sigma_{\rm tot},  \\
 &t\_\sigma_{\rm W}   =  0.98^{+0.08}_{-0.10} \times t\_\sigma_{\rm tot}.
\label{eqAgeUVWtot}
\end{aligned}
\end{equation}
As can be seen, the relative differences between ages from different velocities are $\sim 5\%-10\%$. }
Hereafter, unless otherwise specified, the kinematic age refers to the one derived using the dispersion of total velocity, i.e., $\sigma_{\rm tot}$. 
}

\subsubsection{kinematic ages of planet host stars in  the {\respondtoxiang Galactic} disk}
Besides the element abundances and {\respondtoxiang Galactic} velocities, age is one of the main differences of different {\respondtoxiang Galactic} components.
It has been known that thin disk stars are generally younger than thick disk stars with a dividing age of $\sim 8$ Gyr \citep{1998A&A...338..161F,2013IAUS..292..105H}.
Stars in halo are very old and the age is estimated to be $\sim 10-12$ Gyr \citep{2011A&A...533A..59J,2012Natur.486...90K,2016RAA....16...44G,2019RAA....19....8G}. 
For Hercules stream, it has three substructures. The age distribution is peaked at 4 Gyr and extend to very old age for Hercules a and Hercules b;  and Hercules c
has a more uniform age distribution from $\sim 2-10$ Gyr \citep{2019A&A...629L...6T}.

As mentioned in section \ref{sec.meth.AVR.diffcom}, there is no clear trend between velocity dispersions and ages for stars in Hercules stream and halo.
Thus here we only calculate the kinematic age of planet host stars in {\respondtoxiang Galactic} disks, which contains the majority {\chen (97.9\%)} of our sample. 
This will be useful in future study on the characteristics and evolution of planetary systems related to the {\respondtoxiang Galactic} components and age of host star.

We obtain the kinematic age and uncertainty for stars in thin disk ({\chen 1,894 stars}) and thick disk ({\chen 114 stars}) with the methods described at the beginning of section 4.3.
The typical ages are {\chen $2.84^{+0.32}_{-0.26}$} Gyr, {\chen $10.42^{+1.82}_{-0.73}$} for thin and thick disk respectively.
As shown in Figure \ref{figAgedisk}, the age distribution of stars in the thick disk is generally larger than 8 Gyr, while the thin disk is populated
by younger stars. This is well consistent with previous studies \citep{1998A&A...338..161F,2003A&A...410..527B,2013A&A...560A.109H,2014A&A...562A..71B}.

To explore the relation between $TD/D$ and kinematic age,  we first sorted the plant host sample by the relative probabilities for the  thick-disk-to-thin-disk, i.e., $TD/D$. 
Next, according to $TD/D$, we divided the planet host sample into 10 bins with approximately equal size. 
Then, for each bin, we calculated their kinematic ages. 
As shown in Figure \ref{figAgeTDD}, the kinematic age generally increase with $TD/D$, demonstrating that $TD/D$ is a indicator of age for stars in the {\respondtoxiang Galactic} disk.

\begin{figure*}[!t]
\centering
\includegraphics[width=\textwidth]{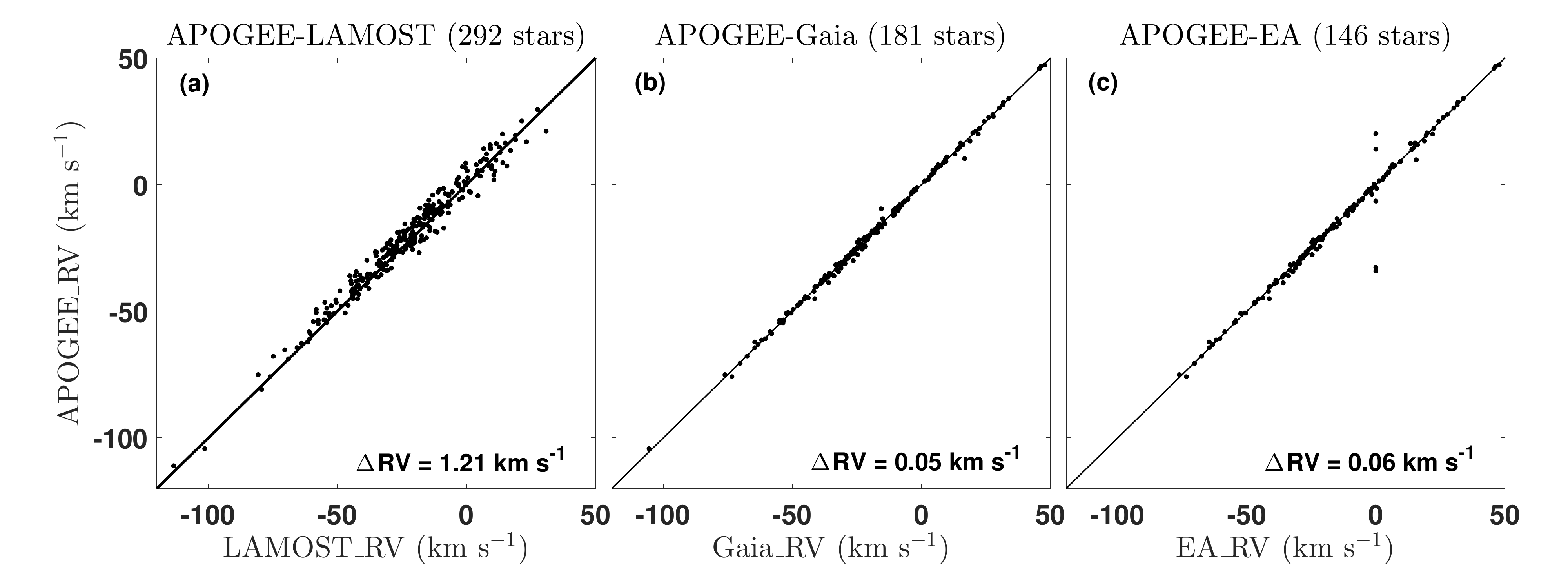}
\caption{\chen The comparison of radial velocities (RV) from different sources: (a). $\rm LAMOST\_RV$ vs. $\rm APOGEE\_RV$; (b). $\rm Gaia\_RV$ vs. $\rm APOGEE\_RV$; (c) $\rm EA\_RV$ vs. $\rm APOGEE\_RV$.
The black lines in the figure indicate where the horizontal and vertical coordinates are equal.
$\Delta RV$ represents the systematic offset in RV comparing to APOGEE measurements.
\label{figRVsources}}
\end{figure*}

\begin{figure}[!t]
\centering
\includegraphics[width=0.5\textwidth]{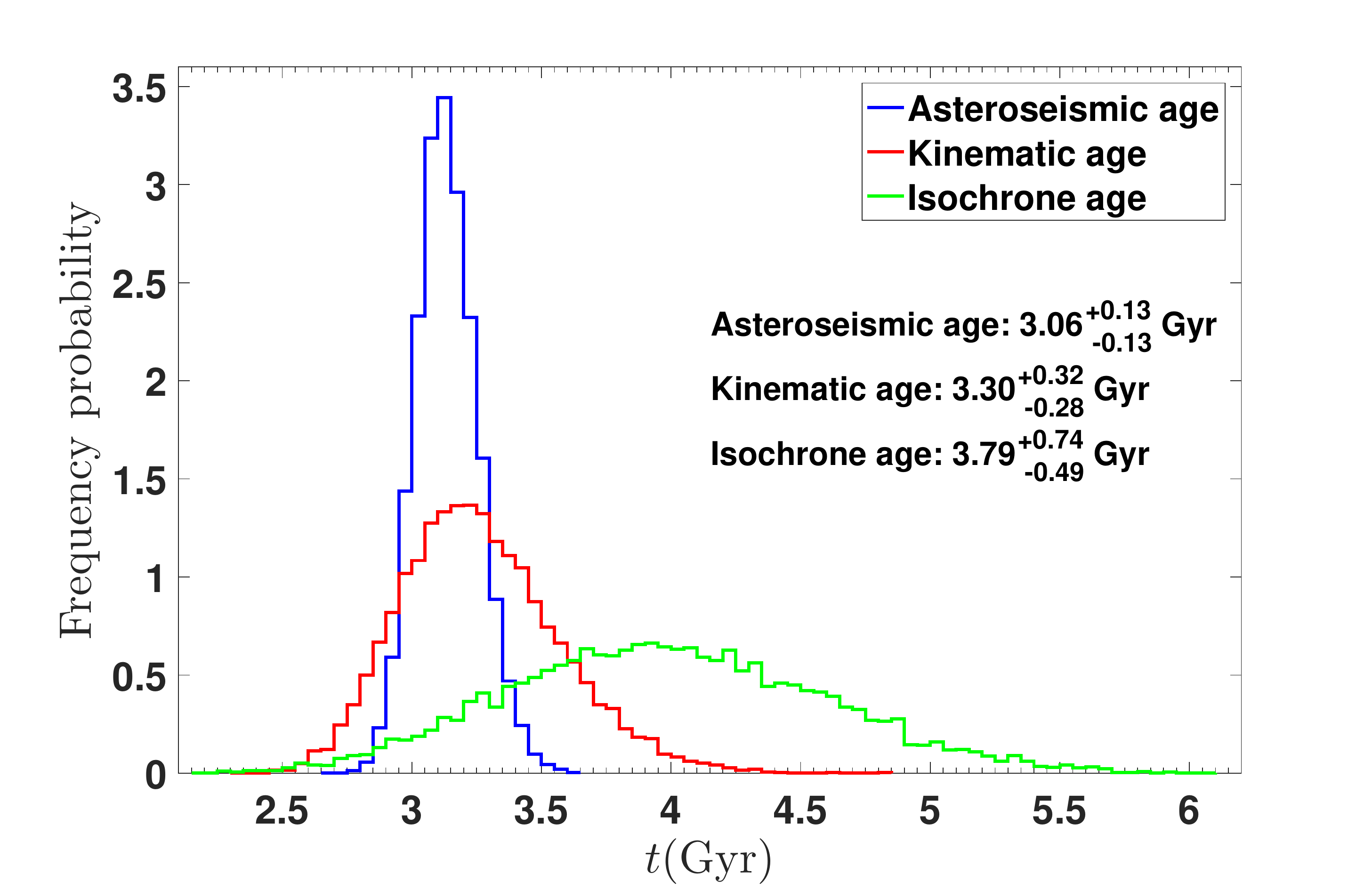}
\caption{Distributions of the asteroseismic age (blue), kinematic age (red) and isochrone age (green) of 54 Kepler stars.
\label{figAgeAK}}
\end{figure}
\section{Discussions}
\label{sec.dis}

\subsection{Systematic Differences in Radial Velocity from Various Sources}
{\respondtohuang As mentioned in section \ref{sec.samp.rv}, we obtained the RV from five sources: APOGEE, LAMOST, RAVE, Gaia, and EA. 
As the RV is one of the basic parameters to calculate the Galactic velocity, the systematic differences in RV will induce differences in the Galactic velocity and then effect the classification of Galactic components. 
Therefore, it is necessary to calibrate the systematic difference in RV from various sources.

Here we choose the APOGEE RV data as standard reference because it has the highest resolution ($R \sim22500$) and thus the most accurate RV measurements.
Then we compare the difference in RV for common stars with measurements from the other sources overlapped with APOGEE : {\chen 292 for LAMOST, 2 for RAVE, 181 for Gaia, and 146 for EA}.
As shown in Figure \ref{figRVsources}, the systematic offsets in RV from APOGEE measurements, $\Delta RV$ are {\chen 1.21, 0.05, 0.06 $\rm km \ s^{-1}$ }for LAMOST, Gaia, and EA respectively.
For RAVE, there is only {\chen 2} common stars and too few to make analysis. Here we refer to \cite{2018AJ....156...90H}, which presents a new catalog of RV standard stars selected from the APOGEE data and find that the systematic offset is only 0.17 $\rm km \ s^{-1}$ for RAVE.
The systematic differences in RV between APOGEE and other sources are all smaller than the typical uncertainties in RV of our sample {\chen 1.58 $\rm km \ s^{-1}$} and thus have no significant influence in the calculation of Galactic velocity and the classification of Galactic components.
}

\subsection{Kinematic Age vs. Asteroseismic Age vs. Isochrone Age}
\label{sec.agecomparison}
{\xie In order to access the reliability of kinematic ages derived in this work, we compare them to ages derived from asteroseismology and isochrone.}
{\xie The Kepler asteroseismic LEGACY sample \citep{2017ApJ...835..173S} provides a well characterized sample of 66 Kepler planet hosts with asteroseismic age estimates whose average uncertainy is $\sim 10\%$.} 
Besides, \cite{2020AJ....159..280B} presents age estimates from isochrone fitting for 186,301 Gaia-Kepler stars with a median uncertainties of 56\%. 
{\xie By cross-matching our planet host catalog (Table \ref{tab:starcatalog}) with the samples in the above two studies, we obtained a common sample of 54 Kepler planet hosts.}
{\xie In Figure \ref{figAgeAK}, we compare the distributions of median asteroseismic age, median ischrone age and kinematic age for the common sample. 
The distributions of median asteroseismic age and median ischrone age were constructed by 10,000 sets of resampled ages based on the reported ages and uncertainties (assuming Gaussian distribution $N(t,\sigma_t)$).
The distribution of kinematic ages were constructed with AVR from Equation \ref{kineage} using a Monte Carlo method which took into account the uncertainties of AVR parameters (Table \ref{tab:AVRpara}) and velocity dispersions.
As can be seen in Figure \ref{figAgeAK}, the results are $3.06^{+0.13}_{-0.13}$ Gyr, $3.30^{+0.32}_{-0.28}$ Gyr and $3.79^{+0.74}_{-0.49}$ Gyr for asteroseismic age, kinematic age, and isochrone age, respectively.
Encouragingly, our kinematic ages match better with the asteroseismic ages, though the three kinds of ages are not inconsistent with each other within their errorbars. }



\subsection{Future Studies}
\label{sec.dis.futurework}
{\jiwei For the clarity and simplicity of the paper, here we apply the revised kinematic methods (section 2 and 3) only to the planet host sample. 
Next, in our second paper of the PAST project (Chen et al. in prep.), we will apply the revised kinematic methods to the whole Kepler star sample, enabling us to further connect stellar kinematic properties to stellar rotations and activities.}

The planet host catalog provides stellar parameters, spatial position, {\respondtoxiang Galactic} velocity and component classification for 2,174 stars, which hosts 2,872 planets. {\xie Furthermore, using AVR as we show in section \ref{sec.kine.age.err}, one can obtain the kinematic age for a group of stars with respective properties. }
{\xie As shown in Figure \ref{figRtheZstars} and Table \ref{tab:starnumber}, these planet hosts spread over different {\respondtoxiang Galactic} components in a wide range of distance up to $~1,500$ pc.}
{\xie With such rich stellar information, future studies are allowed to explore and answer some fundamental questions on  exoplanets, such as,} 
what are the differences in various properties of planetary systems at different positions in the Galaxy with different ages? 
{\jiwei Specifically, in a subsequent paper of the PAST project (Yang et al. in prep), we will study whether/how planetary occurrence and architecture change with the Galactic environment.}
{\xie The answers of these questions will be crucial in constraining various models and theories of planetary formation and evolution.} 

{\chen
\section{Guidelines for Using the Methods and Catalog}
\label{sec.guide}
In this section, we provide the guidelines, cautions and limitations to utilize the revised kinematic methods (section 2 and 3) and {\jiwei planet host catalog (Table \ref{tab:starcatalog})}.

{\jiwei To classify stars into different Galactic components, the key is to calculate the relative probabilities between two different
components (i.e. $TD/D,\ TD/H,\ Herc/D,\ Herc/D$) with Equation (4) and (5), which rely on the X factor and velocity ellipsoid (i.e. $\sigma_U$, $\sigma_V$, $\sigma_W$, and $V_{\rm asym}$)} of different Galactic components. 
Here, we suggest two ways to obtain these kinematic parameters {\jiwei for a given} Galactic position $(R,\ Z)$.
{\jiwei The most easy way is to use our fitting formulae (e.g., Equation \ref{eUVWRZ} with coefficients in Table \ref{tab:eUVWpara} for  $\sigma_U$, $\sigma_V$ and $\sigma_W$, and Equation \ref{eqVasym} for $V_{\rm asym}$). Alternatively, one can conduct interpolation based on our revised characteristics in Table \ref{tab:dispersionsrevised}. 
}

{\jiwei To derive the kinematic age for a group of stars, one can use the AVR (Equation \ref{kineage}) with the revised coefficients in Table \ref{tab:AVRpara}. 
The typical age uncertainty can be estimated from Equation \ref{kineageerr}, which is $\sim10-20\%$ here.}
{\jiwei For the purpose of avoiding potential spacial biases}, we recommend to adopt the total velocity dispersion  (Equation \ref{sigmatot}) {\jiwei when using the AVR (Equation \ref{kineage})}. 

Applying  the  above  revised  methods  to  our  planet host  sample,  we provide a catalog  with kinematic properties (e.g. Galactic position, velocity and $TD/D$)  and  other  basic  parameters (e.g. $T_{\rm eff}$ and element abundances). 
{\jiwei With this catalog,} one can divide planet hosts into bins according to respective properties (e.g., planetary period, multiplicity etc.) and calculate the kinematic age for each bin, which will be practical and useful for statistical {\jiwei studies of age effects on planetary systems.}
{\jiwei Although the kinematic method can not directly measure the ages for individual stars, some of the derived kinematic properties (e.g., $TD/D$) could serve as good age tracers given their significant correlations (e.g., Figure \ref{figAgeTDD}).}

However, there are some notable cautions and limitations, which are listed as follows:

(1). The revised velocity ellipsoid and AVR are strictly applicable for stars within the region calibration sample covers, i.e, within 1.54 kpc to the Sun (corresponding to $R=7.5-10.0$ kpc, $|\theta|<10$ deg $\& \ |Z|=0-1.5$ kpc).
Taking the assumption that the Galactic disk is axis-symmetric \citep[e.g.][]{2014MNRAS.444...62Y,2016MNRAS.462.1697A}, the velocity ellipsoid and AVR will be indenpent of $\theta$ \citep[e.g.][]{2013MNRAS.436..101W}. Thus the criterion $|\theta|<10$ deg are not necessary for the Galactic disk stars.
Besides, with Equation \ref{eUVWRZ} and Table \ref{tab:eUVWpara}, the velocity dispersion can be extrapolated to the region outer to 1.54 kpc.  
However, this extrapolation should be adopted with caution.

(2). Since the additional motions caused  by  binary orbits could affect the stellar kinematic, binaries  should be applied with caution.

(3). Due to the small numbers of Halo and Hercules stream stars in our calibration sample, we adopt the velocity dispersion values derived from stars in the solar neighborhood as in \cite{2014A&A...562A..71B}. 
As the velocity dispersions change with the Galactic position \citep{2013MNRAS.436..101W}, there might be some deviations in the classification of Halo and Hercules stream stars when utilizing the characteristic parameters in Table \ref{tab:dispersionsrevised} and the planet host catalog in Table \ref{tab:starcatalog}.
It will be more reliable to take other parameters \citep[e.g. velocity, element abundance, angular momentum][]{2007ApJ...655L..89B,2011ApJ...738..187L,2017ApJ...845..101B,2019A&A...631A..47K} into consideration.  

(4). There is no clear trend between velocity dispersions and ages for stars belong to Hercules stream and halo in our calibration sample. Therefore, the method to derive kinematic age is only suitable for stars in the Galactic disk.

(5). The kinematic ages and uncertainties derived from Equation \ref{kineage} and \ref{kineageerr} are the typical (median/mean) values for a group of stars.
}

\section{Summary}
\label{sec.sum}
{\xie Since 1995, the discovered exoplanet population has expanded significantly from the solar neighborhood to a much larger area in the Galaxy (Figure \ref{figeudistanceyear}).
We are therefore entering a new era to study exoplanets in a big context of the Galaxy. 
In the {\respondtoxiang Galactic} context, the relations between the properties of planetary systems and the kinematics as well as the ages of planet host stars have yet to be explored.
To answer these questions, we perform a series of studies in a project dubbed as PAST (Planets Across Space and Time).
{\chen In this paper, as the Paper I and the basis of the PAST series, we revisit the kinematic methods for classification of  Galactic components (section \ref{sec.meth.classify}) and estimation of kinematic ages (section \ref{sec.meth.avr}) and apply them to planet host stars (section \ref{sec.res}).} 
}


    For classification of {\respondtoxiang Galactic} components (section \ref{sec.meth.classify}), we adopt the well-used kinematic  approach  as in  \cite{2003A&A...410..527B,2014A&A...562A..71B}. 
    However, so far, the kinematic characteristics of this method has been applied only to the Solar neighborhood within $\sim 100-200$ pc. 
    For this reason, using a calibration sample based on the GAIA and LAMOST data (section \ref{sec.meth.rev.cal}), we extend the kinematic characteristics to $\sim 1,500$ pc (section \ref{sec.meth.rev}, Table \ref{tab:dispersionsrevised}) to cover the majority of planet hosts (Figure \ref{figeudistanceyear} and \ref{figRtheZstars}).

    For estimation of kinematic ages, we refit the Age-Velocity dispersion Relation (AVR) with the calibration sample (section \ref{sec.meth.avr}, Figure \ref{fAVR}). 
    Our AVR is consistent with those in  previous studies (e.g., \cite{2009A&A...501..941H}) but with much smaller internal uncertainties (Table \ref{tab:AVRpara}) thanks to the large and high quality calibration sample.
    Based on this refined AVR, we are able to derive kinematic age with an uncertainty of 10-20\% (section \ref{sec.kine.age.err}), which is a factor of $\sim$3 smaller than those from previous studies (Table \ref{tab:AVRparaHT}).
{\xie Applying the above revised methods to our planet host sample, we then construct a catalog with kinematic properties and other basic parameters for {\chen 2,174} stars (section \ref{sec.res.cat}, Table \ref{tab:starcatalog}) by combining data from Gaia, LAMOST, APOGEE, RAVE and the NASA exoplanet archive (Table \ref{tab:planetnumberprocedure} and section \ref{sec.samp}). 
{The majority (1,894/2,174, 87.1\%) of planet host stars are found to be in the  Galactic thin disk, while  5.2\% (114/2,174) of them belong to the thick disk and only 0.05\% (1/2,174) reside in the halo (Table \ref{tab:starnumber}).} 
As expected, we find that the total velocity, $\rm [\alpha/Fe]$ and the kinematic age generally increase with the relative probabilities for the thick-disk-to-thin-disk, i.e., $TD/D$, while $\rm [Fe/H]$ decreases with $TD/D$ (Figure \ref{figTDDVtotFealpha}, \ref{figAgeTDD}).
The kinematic age is {\chen $2.84^{+0.32}_{-0.26}$} Gyr for the thin disk stars and {\chen $10.42^{+1.82}_{-0.73}$} Gyr for the thick disk stars in the planet host sample (Figure \ref{figAgedisk}).

We also compare our derived kinematic ages with asteroseismic ages and isochrone ages (section \ref{sec.agecomparison}). Our kinematic ages match better with the asteroseismic ages, though the three kinds of ages are generally consistent with each other within their uncertainties (Figure \ref{figAgeAK}).

Future studies of exoplanets in the {\respondtoxiang Galactic} context, e.g., the subsequent papers of our PAST series (section \ref{sec.dis.futurework}), will benefit not only from the catalog of the kinematic properties but also from the revised methods that derived such catalog in this work.
{\chen The important guidelines, cautions and limitation when utilizing our kinematic methods and planet host catalog are described in section \ref{sec.guide}. }

}

\section*{Acknowledgements}
This work is supported by the National Key R\&D Program of China (No. 2019YFA0405100) and the National Natural Science Foundation of China (NSFC; grant No. 11933001, 11973028,  11803012, 11673011,12003027). J.-W.X. also acknowledges the support from the National Youth Talent Support Program and the Distinguish Youth Foundation of Jiangsu Scientific Committee (BK20190005).
H.F.W. is supported by the LAMOST Fellow project, National Key Basic R\&D Program of China via 2019YFA0405500 and funded by China Postdoctoral Science Foundation via grant 2019M653504 and 2020T130563, Yunnan province postdoctoral Directed culture Foundation, and the Cultivation Project for LAMOST Scientific Payoff and Research Achievement of CAMS-CAS.
M. Xiang \& Y. Huang acknowledge the National Natural Science Foundation of China (grant No. 11703035).
C.L. thanks National Key R\&D Program of China No. 2019YFA0405500 and the National Natural Science Foundation of China (NSFC) with grant No. 11835057.

This work has included data from Guoshoujing Telescope.
Guoshoujing Telescope (the Large Sky Area Multi-Object Fiber Spectroscopic Telescope LAMOST) is a National Major Scientific Project built by the Chinese Academy of Sciences. Funding for the project has been provided by the National Development and Reform Commission. LAMOST is operated and managed by the National Astronomical Observatories, Chinese Academy of Sciences.
This work presents results from the European Space Agency (ESA) space mission Gaia. Gaia data are being processed by the Gaia Data Processing and Analysis Consortium (DPAC). Funding for the DPAC is provided by national institutions, in particular the institutions participating in the Gaia MultiLateral Agreement (MLA). The Gaia mission website is https://www.cosmos.esa.int/gaia. 
We acknowledge the NASA Exoplanet archive, which is operated by the California Institute of Technology, under contract with the National Aeronautics and Space Administration under the Exoplanet Exploration Program.
Funding for RAVE has been provided by: the Australian Astronomical Observatory; the Leibniz-Institut fuer Astrophysik Potsdam (AIP); the Australian National University; the Australian Research Council; the French National Research Agency; the German Research Foundation (SPP 1177 and SFB 881); the European Research Council (ERC-StG 240271 {\respondtoxiang Galactic}a); the Istituto Nazionale di Astrofisica at Padova; The Johns Hopkins University; the National Science Foundation of the USA (AST-0908326); the W. M. Keck foundation; the Macquarie University; the Netherlands Research School for Astronomy; the Natural Sciences and Engineering Research Council of Canada; the Slovenian Research Agency; the Swiss National Science Foundation; the Science \& Technology Facilities Council of the UK; Opticon; Strasbourg Observatory; and the Universities of Groningen, Heidelberg and Sydney.
The RAVE web site is at https://www.rave-survey.org.
Funding for the Sloan Digital Sky Survey IV has been provided by the Alfred P. Sloan Foundation, the U.S. Department of Energy Ofﬁce of Science, and the Participating Institutions. The SDSS website is www.sdss.org.

\bibliography{library.bib}

\newpage

\clearpage
\begin{table*}
\centering
\caption{Revised characteristics at different Galactic radii ($R$) and heights ($Z$) for different Galactic components using the calibration sample.} 
{\footnotesize
\label{tab:dispersionsrevised} 
\begin{tabular}{c|c|cccccccccccc} \hline
$|Z|$ & $R$ & $\sigma^{\rm D}_{\rm U}$ & $\sigma^{\rm D}_{\rm V}$ & $\sigma^{\rm D}_{\rm W}$ & $V^{\rm D}_{\rm asym}$ & $\sigma^{\rm TD}_{\rm U}$ & $\sigma^{\rm TD}_{\rm V}$ & $\sigma^{\rm TD}_{\rm W}$ & $V^{\rm TD}_{\rm asym}$ & \multirow{2}{*}{$X_{\rm D}$} & \multirow{2}{*}{$X_{\rm TD}$} & \multirow{2}{*}{$X_{\rm H}$} & \multirow{2}{*}{$X_{\rm Herc}$}\\ 
(kpc) & (kpc) & \multicolumn{8}{c}{---------------------------~~($\rm km \ s^{-1}$)~~---------------------------}  & & & & \\ \hline 

\multirow{6}{*}{$0-0.1$}       & $7.5-8.0$  & 37 & 22 & 15 & $-16$ & 63  & 37& 31 & -41& 0.81& 0.13 & 0.0010&0.06\\
                                             & $8.0-8.5$  & 34 & 21 & 16 & $-14$ & 65  & 39 & 35 & $-44$& 0.84&0.10&0.0013&0.06\\
                                             & $8.5-9.0$  & 35 & 20 & 16 & $-15$ & 70  & 37 & 33 & $-51$& 0.85 &0.10&0.0013&0.05\\
                                             & $9.0-9.5$  & 33 & 19 & 15 & $-12$ & 70  & 35 & 34 & $-51$& 0.87&0.09&0.0014&0.04\\
                                             & $9.5-10.0$ & 33 & 19 & 16 & $-11$ & 68 & 37 & 31 & $-48$& 0.89&0.08&0.0016&0.03\\ \hline
 \multirow{6}{*}{$0.1-0.2$}    & $7.5-8.0$  & 32 & 23 & 15 & $-13$ & 69  &  37  & 39 & $-43$&0.78&0.16&0.0011&0.06\\
                                              & $8.0-8.5$  & 35  & 22 & 17 & $-15$ & 66  & 42 & 36  & -45&0.79&0.14&0.0010&0.07\\
                                              & $8.5-9.0$  & 35  & 21 & 16 & $-15$ & 71  & 41 & 36 & -52&0.81&0.13&0.0014&0.06\\
                                              & $9.0-9.5$  & 32  & 20 & 16 & $-13$ & 71  & 42 & 34 & $-52$&0.85&0.12&0.0012&0.04\\
                                              & $9.5-10.0$ & 30 & 20 & 16 & $-11$ & 70 & 40 & 34 &$-50$&0.87&0.10&0.0019&0.03\\ \hline
\multirow{6}{*}{$0.2-0.3$}    & $7.5-8.0$  & 36 & 22 & 16 & $-16$ & 67  & 39 & 37 & $-46$&0.75&0.20&0.0016&0.07\\
                                             & $8.0-8.5$  & 37 & 23 & 17 & $-17$ & 68  & 42 & 40 & $-47$&0.76&0.17&0.0015&0.07\\
                                             & $8.5-9.0$  & 37 & 22 & 18 & $-17$ & 72  & 42 & 37 &$-54$&0.78&0.16&0.0017&0.06\\
                                             & $9.0-9.5$  & 34 & 21 & 16 & $-14$ & 71  & 40 & 36 &$-52$&0.81&0.14&0.0016&0.05\\
                                             & $9.5-10.0$ & 31 & 21 & 16 & $-11$& 67  & 39 & 35 &$-48$&0.84&0.13&0.0140&0.03 \\ \hline
\multirow{6}{*}{$0.3-0.4$}     & $7.5-8.0$  & 38 & 24 & 19 & $-18$ & 68  & 41  & 38 &-47&0.71&0.24&0.0016&0.05\\
                                              & $8.0-8.5$  & 40 & 23 & 18 & $-19$ & 68  & 41 & 41 &$-47$&0.71&0.21&0.0020&0.08\\
                                              & $8.5-9.0$  & 38 & 22 & 18 & $-17$ & 71  & 41 & 38 &$-52$&0.75&0.19&0.0020&0.06\\
                                              & $9.0-9.5$  & 36 & 21 & 17 & $-15$ & 71  & 42 & 37 &$-52$&0.77&0.17&0.0020&0.06\\
                                              & $9.5-10.0$ & 34 & 21 & 16 & $-14$ & 71 & 41 & 35 &$-52$&0.80&0.16&0.0025&0.04\\ \hline
\multirow{6}{*}{$0.4-0.55$}     & $7.5-8.0$  & 42 & 26 & 18 & $-21$ & 71  & 42  & 40 &-52&0.65&0.29&0.0021&0.06\\
                                              & $8.0-8.5$  & 41 & 24 & 19 & $-21$ & 70  & 41 & 42 &$-50$&0.66&0.26&0.0026&0.08\\
                                              & $8.5-9.0$  & 39 & 23 & 18 & $-19$ & 71  & 42 & 38 &$-52$&0.70&0.23&0.0022&0.07\\
                                              & $9.0-9.5$  & 37 & 22 & 16 & $-17$ & 71  & 41 & 38 &$-52$&0.73&0.21&0.0024&0.06\\
                                              & $9.5-10.0$ & 37 & 22 & 16 & $-16$ & 70 & 40 & 35 &$-50$&0.75&0.19&0.0025&0.06\\ \hline
\multirow{6}{*}{$0.55-0.75$}     & $7.5-8.0$  & 42 & 27 & 21 & $-21$ & 71  & 45  & 41 &-52&0.55&0.37&0.0042&0.08\\
                                              & $8.0-8.5$  & 44 & 24 & 20 & $-23$  & 72  & 43 & 42 &$-54$&0.58&0.34&0.0035&0.08\\
                                              & $8.5-9.0$  & 40 & 24 & 19 & $-20$ & 72 & 43 & 40 &$-54$&0.61&0.31&0.0026&0.08\\
                                              & $9.0-9.5$  & 38 & 23 & 19 & $-18$ & 72  & 43 & 40 &$-52$&0.65&0.29&0.0026&0.06\\
                                              & $9.5-10.0$ & 38 & 23 & 18 & $-16$ & 71 & 40 & 40 &$-52$&0.68&0.26&0.0029&0.06\\ \hline
\multirow{6}{*}{$0.75-1.0$}     & $7.5-8.0$  & 45 & 29 & 23 & $-24$ & 71  & 43  & 44 &-52&0.43&0.48&0.0055&0.08\\
                                              & $8.0-8.5$  & 45 & 26 & 21 & $-22$ & 73  & 43 & 43 &$-55$&0.46&0.45&0.0053&0.08\\
                                              & $8.5-9.0$  & 42 & 23 & 20 & $-21$ & 73  & 45 & 41 &$-55$&0.50&0.42&0.0039&0.09\\
                                              & $9.0-9.5$  & 39 & 23 & 20 & $-19$ & 72  & 43 & 40 &$-54$&0.56&0.36&0.0026&0.08\\ \hline

\multirow{6}{*}{$1.0-1.5$}     & $7.5-8.0$  & 46 & 31 & 25 & $-25$ & 72  & 45  & 46 &-54&0.25&0.63&0.0129&0.10\\
                                              & $8.0-8.5$  & 45& 28 & 23 & $-22$ & 73  & 45 & 44 &$-55$&0.28&0.61&0.0100&0.10\\
                                              & $8.5-9.0$  & 42 & 36 & 22 & $-24$ & 73  & 45 & 43 &$-55$&0.31&0.59&0.0103&0.09\\
                                              & $9.0-9.5$  & 43 & 25 & 21 & $-23$ & 74  & 45 & 40 &$-57$&0.34&0.56&0.0056&0.09\\ \hline \hline

\end{tabular}}
\end{table*}

\newpage


\begin{table*}[!h]
\centering
\caption{The catalogue of kinematic properties and other basic properties for the combined planet host stars}
{\footnotesize
\label{tab:starcatalog}
\begin{tabular}{p{1cm} p{2.5cm} p{1.25cm} p{1.25cm} p{8.75cm}} \hline
   Column & Name & Format & Units & description\\\hline \hline
\multicolumn{5}{c}{Parameters obtained from Gaia, APOGEE, RAVE, LAMOST and NASA exoplanet archive (EA)} \\ \hline
  1 & Gaia\_ID & Long & & Unique Gaia source identifier \\
  2 & LAMOST\_ID & string & & LAMOST unique spectral ID\\
  3 & {APOGEE}\_ID & string & & APOGEE unique spectral ID\\  
  4 & {RAVE}\_ID & string & & RAVE unique spectral ID\\
  5 & {pl\_hostname} & string & & NASA Exoplanet archive unique planet host name\\
  6 & {Kepler}\_ID & integer & & Kepler Input Catalog (KIC) ID\\
  7 & {Gaia} RA & Double & deg & Barycentric right ascension\\  
  8 & {Gaia} Dec & Double & deg & Barycentric Declination\\ 
  9 & {Gaia} parallax & Double & mas & Absolute stellar parallax\\
 10 & {Gaia} e\_parallax & Double & mas & Standard error of parallax\\
 11 & {Gaia} pmra & Double & mas $\rm yr^{-1}$ & Proper motion in right ascension direction \\
 12 & {Gaia} e\_pmra & Double & mas $\rm yr^{-1}$ & Standard error of proper motion in right ascension direction\\
 13 & {Gaia} pmdec & Double & mas $\rm yr^{-1}$ &Proper motion in declination direction \\
 14 & {Gaia} e\_pmdec & Double & mas $\rm yr^{-1}$ & Standard error of proper motion in declination direction \\
 15 & {Gaia} G mag & Double & mag & \textit{Gaia} \textit{G} band apparent magnitude\\  
 16 & $T_{\rm eff}$ & Float & K & Effective temperature from RAVE, LAMOST, APOGEE, Gaia, EA\\
 17 & ${\rm flag}\_T_{\rm eff}^1$ & integer &  & flag represents which source each value is collected from \\
 18 & $\log g$ & Float &  & Surface gravity from RAVE, LAMOST, APOGEE, Gaia, EA\\
 19 & ${\rm flag}\_\log g^1$ & integer &  & flag represents which source each value is collected from \\
 20 & $\rm [Fe/H]$ & Float & dex &  Metallicity from RAVE, LAMOST, APOGEE, Gaia, EA\\
 21 & ${\rm flag}\_{\rm [Fe/H]}^1$ & integer &  & flag represents which source each value is collected from \\
 22 & $\rm [\alpha/Fe]$ & Float & dex & $\rm \alpha$ elements abundance from RAVE, LAMOST, APOGEE, Gaia, EA \\ 
 23 & ${\rm flag}\_{\rm [\alpha/Fe]}^1$ & integer &  &flag represents which source each value is collected from \\
 24 & rv  &  Double & km $\rm s^{-1}$ &  Radial velocity from APOGEE, RAVE, Gaia, LAMOST, EA\\          
 25 & e\_rv  &  Double & km $\rm s^{-1}$ & Error of radial velocity\\ 
 26 & ${\rm flag}\_{\rm rv}^1$ & integer &  & flag represents which source each value is collected from \\ \hline \hline
 \multicolumn{5}{c}{Parameters derived in this work} \\ \hline
 27 & $R$ & Double & kpc & Galactocentric Cylindrical radial distance \\  
 28 & $\theta$ & Double & deg & Galactocentric Cylindrical azimuth angle \\
 29 & $Z$ & Double & kpc & Galactocentric Cylindrical vertical height \\
 30 & $V_{R}$ & Double & km $\rm s^{-1}$ & Galactocentric Cylindrical $R$ velocities  \\ 
 31 & $V_{\theta}$ & Double & km $\rm s^{-1}$ & Galactocentric Cylindrical $\theta$ velocities  \\ 
 32 & $V_{Z}$ & Double & km $\rm s^{-1}$ & Galactocentric Cylindrical $Z$ velocities  \\ 
 33 & $U_{\rm LSR}$ & Double & km $\rm s^{-1}$ & Cartesian Galactocentric $X$ velocity to the LSR\\ 
 34 & e\_$U_{\rm LSR}$ & Double & km $\rm s^{-1}$ & error of Cartesian Galactocentric $X$ velocity to the LSR\\ 
 35 & $V_{\rm LSR}$ & Double & km $\rm s^{-1}$ & Cartesian Galactocentric $Y$ velocity to the LSR\\ 
 36 & e\_$V_{\rm LSR}$ & Double & km $\rm s^{-1}$ & error of Cartesian Galactocentric $Y$ velocity to the LSR\\ 
 37 & $W_{\rm LSR}$ & Double & km $\rm s^{-1}$ & Cartesian Galactocentric $Z$ velocity to the LSR\\ 
 38 & e\_$W_{\rm LSR}$ & Double & km $\rm s^{-1}$ & error of Cartesian Galactocentric $Z$ velocity to the LSR\\ 
 39 & $TD/D$ & Double & & Thick disc to thin disc membership probability ratio \\
 40 & $TD/H$ & Double & & hick disc to halo membership probability ratio  \\
 41 & $Her/D$ & Double & & Hercules stream to thin disc membership probability ratio  \\
 42 & $Herc/TD$ & Double & & Hercules stream to thick disk membership probability ratio \\ \hline \hline
\end{tabular}}
\flushleft
{\scriptsize
Note 1: The flag represents which source each value is collected from: 1 for APOGEE \citep{2020ApJS..249....3A}; 2 for RAVE \citep{2017AJ....153...75K}; 3 for Gaia \citep{2018A&A...616A...1G,2018A&A...616A..11G}; 4 for LAMOST \citep{2017MNRAS.467.1890X}; 5 for NASA exoplanet archive (https://exoplanetarchive.ipac.caltech.edu/); 0 for not available.
\\}
\end{table*}

\clearpage

\end{document}